\documentclass[12pt]{article}
\pdfoutput=1
\usepackage{amssymb,amsmath,mathrsfs,bm,feynmf,setspace} 
\usepackage{graphicx}
\usepackage{adjustbox}
\usepackage[tight]{subfigure} 
\usepackage{framed}
\usepackage{epsfig} 
\usepackage{epstopdf}
\usepackage{latexsym}
\usepackage{psfrag}   
\usepackage{subfigure}  
\usepackage{booktabs}  
\usepackage{braket}
\usepackage{slashed}
\usepackage[numbers,sort&compress]{natbib}
\usepackage[toc]{appendix}
\usepackage{color,soul} 
\usepackage{datetime}
\usepackage[
      colorlinks=true, 
      linkcolor=darkblue,  
      urlcolor=blue,    
      filecolor=blue,     
      citecolor=darkgreen, 
      linktocpage=true,
      pdfstartview=FitV,
      bookmarksopen=true    
      ]{hyperref}

\usepackage{float}
\usepackage[T1]{fontenc}

\definecolor{darkblue}{rgb}{0.2, 0, 0.8}
\definecolor{darkgreen}{rgb}{0.2, 0.71, 0}

\numberwithin{equation}{section}

\newcommand{\req}[1]{(\ref{#1})} 
\newcommand{\labell}[1]{\label{#1}}

\newcommand{\bea}{\begin{eqnarray}}
\newcommand{\eea}{\end{eqnarray}}
\newcommand{\ba}{\begin{eqnarray}}
\newcommand{\ea}{\end{eqnarray}}

\newcommand{\beq}{\begin{equation}}
\newcommand{\eeq}{\end{equation} }
\newcommand{\beqa}{\begin{eqnarray}}
\newcommand{\eeqa}{\end{eqnarray}}
\newcommand{\beqar}{\begin{eqnarray*}}
\newcommand{\eeqar}{\end{eqnarray*}}

\newcommand{\ssc}{\scriptscriptstyle}
\newcommand{\eg}{{\it e.g.,}\ }
\newcommand{\ie}{{\it i.e.,}\ }

\newcommand{\ctt}{C_{\ssc T}}

\renewcommand{\href}[2]{#2}

\begin{document}  


\begin{titlepage}

\vspace*{2.0cm}

\begin{center}
{\LARGE \bf Comments on Squashed-sphere\\ \vspace{0.2cm} Partition Functions}\\

\vspace*{1.3cm}

{\bf Nikolay Bobev, Pablo Bueno and Yannick Vreys}

\vspace*{.75cm}

Instituut voor Theoretische Fysica, KU Leuven,\\ Celestijnenlaan 200D, B-3001 Leuven, Belgium

\vspace*{.75cm}

\tt{nikolay.bobev@kuleuven.be, pablo@itf.fys.kuleuven.be, yannick.vreys@fys.kuleuven.be}

\end{center}


\begin{abstract}  

\vspace*{0.2cm}

\noindent   We study the partition function of odd-dimensional conformal field theories placed on spheres with a squashed metric. We establish that the round sphere provides a local extremum for the free energy which, in general, is not a global extremum. In addition, we show that the leading quadratic correction to the free energy around this extremum is proportional to the coefficient, $C_{ \ssc T}$, determining the two-point function of the energy-momentum tensor in the CFT. For three-dimensional CFTs, we compute explicitly this proportionality constant for a class of squashing deformations which preserve an $SU(2)\times U(1)$ isometry group on the sphere. In addition, we evaluate the free energy as a function of the squashing parameter for theories of free bosons, free fermions, as well as CFTs holographically dual to Einstein gravity with a negative cosmological constant. We observe that, after suitable normalization, the dependence of the free energy on the squashing parameter for all these theories is nearly universal for a large region of parameter space and is well approximated by a simple quadratic function arising from holography. We generalize our results to five-dimensional CFTs and, in this context, we also study theories holographically dual to six-dimensional Gauss-Bonnet gravity.

\end{abstract}

\bigskip

\end{titlepage}

\setcounter{tocdepth}{2}
{\small
\setlength\parskip{-0.5mm} 
\tableofcontents
}

\section{Introduction and summary of results} 
\label{sec:intro} 

Coupling a quantum field theory to background fields is a powerful tool which can be utilized to gain insights into the dynamics of the theory. This idea has been used often in studying quantum field theories, and is typically most powerful when the QFT at hand enjoys extra symmetries which constrain its dynamics. A well-known example that has gathered a lot of attention recently is coupling supersymmetric theories in various dimensions to background metrics and background gauge fields for the global symmetries of the theory. One can then often rely on the powerful tool of supersymmetric localization to extract many exact results for large classes of QFTs --- see \cite{Pestun:2016zxk} for a review. In the absence of supersymmetry, one loses much of the technical control but, as we show below, if the theory at hand possesses conformal symmetry, some progress can still be made.

Our interest here is in studying Euclidean conformal field theories in odd dimensions, and couple them to a curved background metric. Every CFT has a conserved energy momentum tensor which naturally couples to the background metric and, thus, the type of deformations we study have a universal character and exist for any CFT. We choose to study theories in odd dimensions since these are not plagued by conformal anomalies. In addition, there is a natural class of deformations of the Einstein metric on odd-dimensional spheres which allow for many explicit computations. 

The main focus of our work is to understand how the partition function or, equivalently, the free energy of general CFTs depends on the background metric. To achieve this, we can place the CFT on a conformally flat background metric $\bar{g}_{\mu\nu}$, a prime example being the round $d$-dimensional sphere $S^d$, and switch on a ``squashing'' deformation of this metric controlled by the symmetric tensor $\epsilon h_{\mu\nu}$ where $\epsilon$ is a real parameter. We can then study the free energy\footnote{In this work we focus on the real part of the free energy of odd-dimensional CFTs and thus avoid the subtleties associated with contact terms that can affect its imaginary part, see \cite{Closset:2012vg,Closset:2012vp}} of the CFT as a function of $\epsilon$. In Section \ref{party}, we show that, for any odd-dimensional CFT, this function has the following series expansion for small $\epsilon$
\begin{equation}\label{Fintro}
\mathcal{F}(\epsilon)=\mathcal{F}(0)+\frac{\epsilon^2}{2}\mathcal{F}^{\prime\prime}(0)+\mathcal{O}(\epsilon^3)\, ,
\end{equation}
where
\begin{equation}\label{F2primeintro}
\mathcal{F}^{\prime\prime}(0)=\tilde{G}(\bar{g}_{\mu\nu},h_{\mu\nu},d)\, \ctt  \, ,
\end{equation}
and $\tilde{G}(\bar{g}_{\mu\nu},h_{\mu\nu},d)$ is a function which depends (see \req{conj2t} below) on the corresponding conformally flat metric $\bar{g}_{\mu\nu}$, the metric deformation $h_{\mu\nu}$ and the spacetime dimension, $d$.\footnote{A similar formula has been derived for entanglement entropies of deformed spherical and planar regions. In Section \ref{EESD} we provide some comments on the connection between these two results.} The real number $\ctt$ is theory dependent and positive for unitary CFTs. It is often referred to as the ``central charge'' and determines the two-point function of the energy-momentum tensor. In particular, for CFTs in Euclidean flat space $\mathbb{R}^d$ with coordinates $X_a$ one has \cite{Osborn}\footnote{\label{foo}Note that our convention for $\ctt$ is the same as in \cite{Bobev:2015jxa}, and differs from that in \eg \cite{Osborn,Buchel:2009sk} by a factor of $1/\mathcal{S}_d^2=\Gamma[d/2]^2/(4\pi^d)$. For example, in our convention, $\ctt=d/(d-1)$ for a free real scalar field.}
\begin{align} \label{t2p}
\braket{ T_{ab} (X)\, T_{cd}(0) }_{\mathbb{R}^d}=\frac{1}{\mathcal{S}^2_{d-1}}\frac{\ctt}{|X|^{2d}}\,\mathcal{I}_{ab,cd}(X)\, ,
\end{align} 
where $\mathcal{S}_{d}\equiv2\pi^{d/2}/\Gamma[d/2]$ is the area of the unit $(d-1)$-sphere and the tensorial object $\mathcal{I}_{ab,cd}(X)$ reads
\begin{equation}\label{Iabcddef}
\mathcal{I}_{ab,cd}(X)\equiv\frac{1}{2}\left(I_{ac}(X)I_{bd}(X)+I_{ad}(X)I_{bc}(X)\right)-\frac{1}{d}\delta_{ab}\delta_{cd}\, , \quad \text{where} \quad I_{ac}(X)\equiv\delta_{ac}-2\frac{X_aX_c}{|X|^2}\, .
\end{equation}
Computing the explicit form of $\mathcal{F}^{\prime\prime}(0)$ and the higher order terms in \eqref{Fintro} is hard for general squashing deformations. To make further progress, we focus on squashed spheres $S^{d}_{\epsilon}$, which are characterized by simple explicit metrics with a large isometry group. The squashed spheres we study are singled out by the fact that, just like the usual odd-dimensional round spheres $S^d$, they can be described as ``Hopf fibrations'' over the complex projective spaces $\mathbb{CP}^{k}$ (we use the notation $k\equiv (d-1)/2$), \ie
\begin{equation}\label{fib}
S^1\hookrightarrow S_{\epsilon}^d\rightarrow \mathbb{CP}^{k}\,.
\end{equation}
The metric we put on these squashed spheres is given by 
\begin{equation}\label{squashedSd}
\mathrm{d}s_{S^{d}_{\alpha}}^2 =\frac{1}{(d+1)} \mathrm{d}s^2_{\mathbb{CP}^{k}}+\frac{1}{(1+\alpha)}\left(\mathrm{d}\psi+\frac{A_{\mathbb{CP}^{k}}}{(d+1)}\right)^2\;,
\end{equation}
where $\mathrm{d}s^2_{\mathbb{CP}^{k}}$ is the Einstein metric on the complex projective space $\mathbb{CP}^{k}$ normalized in a way such that $R_{ab}=g_{ab}$, $\psi$ is a periodic coordinate parametrizing the $S^1$ and $J=\mathrm{d}A_{\mathbb{CP}^{k}}$, where $J$ is the K\"ahler form on $\mathbb{CP}^{\frac{d-1}{2}}$ --- see Appendix \ref{TNn} for more details. Note that we find it convenient to parametrize the squashing deformation in \eqref{Fintro} as $\epsilon = -\frac{\alpha}{1+\alpha}$ with $\alpha\geq -1$. The metric on the usual round sphere is restored for $\alpha=\epsilon=0$. For non-vanishing $\alpha$, the metric in \eqref{squashedSd} preserves an $SU(\frac{d+1}{2})\times U(1)$ subgroup of the $SO(d+1)$ isometry group of the round $d$-dimensional sphere.

The free energy of an odd-dimensional CFT on the round sphere is a useful quantity since it is conjectured to be monotonic along unitary RG flows. More precisely the quantity conjectured to decrease along RG flows for odd-dimensional QFTs is \cite{Klebanov:2011gs} 
\begin{equation}\label{Ftildedef}
\tilde{\mathcal{F}} = (-1)^{(d-1)/2}\log |Z_{S^d}|\;.
\end{equation}
This $F$-theorem was proven for three-dimensional CFTs in \cite{Casini:2012ei} using entanglement entropy techniques (see \cite{Pufu:2016zxm} for a review and further references), but remains a conjecture for higher-dimensional theories. The squashing deformations of the round sphere metric that we study here, break conformal invariance in the CFT and thus induce an RG flow. This flow is however outside the scope of the established $F$-theorem, since the induced CFT perturbation is driven by the energy-momentum tensor, which is an operator of spin 2. Nevertheless, it is still natural to expect that the free energy should be a local maximum in the space of deformations controlled by the parameter $\epsilon$ in \eqref{Fintro}. This expectation indeed bears out. For CFTs in three and five dimensions, and the particular metric deformation in \eqref{squashedSd},  we are able to explicitly compute the quadratic correction to $\tilde{\mathcal{F}}$ in the small $\epsilon$ expansion, and find\footnote{Note that throughout the paper, we use the notation $\mathcal{F}^{(n)}\equiv d^n \mathcal{F}/d\epsilon^n$, \ie the derivatives are taken with respect to $\epsilon$, not $\alpha$.}
\begin{equation}\label{introFprresults}
\tilde{\mathcal{F}}^{\prime\prime}_{S^{3}_{\alpha}}(0)= - \frac{\pi^2}{48}\ctt\,, \quad \quad
\tilde{\mathcal{F}}^{\prime\prime}_{S^{5}_{\alpha}}(0)= - \frac{3\pi^2}{320}\ctt\;.
\end{equation}
For unitary CFTs, the central charge $\ctt$ is positive and thus the free energy on the round sphere $\tilde{\mathcal{F}}(0)$ is a local maximum in the space of squashing deformations given by \eqref{squashedSd}.

In addition to the general results discussed above, we are able to compute the function $\tilde{\mathcal{F}}(\epsilon)$ for general values of the squashing parameter in various theories in three and five dimensions. In three-dimensional CFTs, we can compute this free energy for a conformally coupled scalar and a free Dirac fermion by utilizing the explicitly known eigenvalues of the Laplacian and the Dirac operator for the metric in \eqref{squashedSd}. The calculation then boils down to an evaluation of a Gaussian path integral which we are able to carry out numerically using an approach similar to the one employed in \cite{Anninos:2012ft, Anninos:2013rza,Bobev:2016sap}. These explicit calculations are in harmony with the general results in \eqref{Fintro} and \eqref{introFprresults}. For three-dimensional CFTs with weakly coupled gravity duals, we have access to analytic results for $\tilde{\mathcal{F}}(\epsilon)$. The CFT at hand placed on the background with metric \eqref{squashedSd} is described holographically by the well-known AdS-Taub-NUT solution \cite{Hawking:1998ct,Chamblin:1998pz}. Via the standard holographic dictionary, the partition function of the CFT is captured by the properly regularized on-shell gravitational action for such solution. The result is quite simple and somewhat surprising. The cubic and higher order terms in $\epsilon$ in \eqref{Fintro} vanish, and $\ctt$ completely controls the simple quadratic dependence of the squashed sphere partition function on $\epsilon$ through the expressions in \eqref{Fintro}, \eqref{F2primeintro}, and \eqref{introFprresults}. We should stress that the free energy for CFTs in this squashed sphere background has been studied before in \cite{Hartnoll:2005yc}, where the free energy of the free and interacting $O(N)$ model was discussed in the context of higher-spin holography --- see also \cite{Anninos:2012ft, Anninos:2013rza,Bobev:2016sap} for a more recent discussion. It is also worth pointing out that the partition function of three-dimensional $\mathcal{N}=2$ SCFTs on the squashed sphere with the metric \eqref{squashedSd} was studied in a number of papers using the tools of supersymmetric localization --- see for example \cite{Hama:2011ea,Imamura:2011wg}. In the supersymmetric context, in addition to the background metric, one has to turn on a non-trivial background gauge field for the $U(1)$ R-symmetry current present in every such CFT. Due to this reason, the results in \cite{Closset:2012ru}, although similar in spirit, are different from ours since we do not use supersymmetry and the only background field in our setup is the metric.

To confirm the general structure of the free energy, and the result in \eqref{introFprresults}, we also evaluate explicitly the function $\tilde{\mathcal{F}}(\epsilon)$ in five dimensions for two particular classes of CFTs. For a five-dimensional conformally coupled scalar, we can again explore the high-symmetry of the metric in \eqref{squashedSd} to evaluate explicitly the eigenvalues of the Laplacian and subsequently study numerically the resulting path integral. We find excellent agreement with the general formulas in \eqref{Fintro}, \eqref{F2primeintro}, and \eqref{introFprresults}. In addition to this free CFT, we also employ a bottom-up holographic approach and use Gauss-Bonnet gravity with a negative cosmological constant as a dual description to a large class of putative five-dimensional CFTs. This class of gravitational theories admit AdS-Taub-NUT type solutions which have the squashed metric in \eqref{squashedSd} as an asymptotic boundary. Using holographic technology, we are able to compute explicitly the squashed sphere free energy, and find that it is a cubic function of $\epsilon$, given in \eqref{tnfgb} below, with coefficients determined by the central charge $\ctt$ and one of the coefficients, usually dubbed $t_2$, determining the three-point function of the energy-momentum tensor. 

The outline of this paper is as follows. We begin in the next Section with a general discussion of the partition function of odd-dimensional CFTs on deformed conformally flat manifolds. In Section \ref{sec:3}, we study the partition function of CFTs on a squashed three sphere and compute this function explicitly for free boson and fermion theories as well as for CFTs with a weakly coupled Einstein holographic dual. In Section \ref{sec:5}, we extend this analysis to the case of five-dimensional CFTs placed on a squashed five sphere and compute the partition function explicitly for a free boson as well as for CFTs dual to six-dimensional Gauss-Bonnet gravity. We conclude in Section \ref{sec:disc} with some comments on supersymmetric theories on the squashed sphere as well as on an extension of our results to more general deformations of the round sphere. We also discuss the connection between our study and recent results on the shape-dependence of entanglement entropy in CFTs. In the four appendices, we summarize the structure of three-point correlation functions of the energy-momentum tensor in CFTs, present details on the holographic evaluation of the partition function, the spectrum of the scalar Laplacian on squashed spheres, as well as a summary of the numerical approach to calculating the partition function of free theories.

\section{Partition functions on deformed manifolds}
\label{party}

Let us consider the Euclidean partition function of a general CFT on a $d$-dimensional\footnote{We will soon restrict $d$ to be an odd integer.} manifold $\mathcal{M}$ with metric $g_{\mu\nu}$
\begin{align}
	Z=\int \mathcal{D} \varphi\, e^{-I[\varphi,g_{\mu\nu}]}\, ,
\end{align}
where $\varphi$ stands schematically for the set of dynamical fields in the theory. We wish to understand how the free energy $\mathcal{F}\equiv-\log Z$ changes under small deformations of the metric. Let us parametrize such deformations as
\begin{equation}\label{pepe}
g_{\mu\nu}= \bar{g}_{\mu\nu}+\epsilon\, h_{\mu\nu}\, ,
\end{equation}
where $\bar{g}_{\mu\nu}$ is some arbitrary reference metric on $\mathcal{M}$, $\epsilon$ is a real constant, and $h_{\mu\nu}$ encodes the geometry of the perturbation. If we assume that $|\epsilon|\ll1$, we can expand the action in a power series in $\epsilon$ as follows 
\begin{align}\notag
I[g_{\mu\nu}]&=I[\bar{g}_{\mu\nu}]-\frac{\epsilon}{2} \int d^dx \sqrt{\bar g(x)} \,h^{\mu\nu}(x) T_{\mu\nu}(x)  \\ \notag &-\frac{\epsilon^2}{4} \int d^dx  \sqrt{\bar g(x)} \left[\frac{h(x)}{2}h^{\mu\nu}(x) T_{\mu\nu}(x)+  \int d^dy \sqrt{\bar g(y)} \, h^{\mu\nu}(x)h^{\rho\sigma}(y) \frac{\delta T_{\mu\nu}(x)}{\sqrt{\bar{g}(y)}\delta g^{\rho \sigma}(y)} \right]\\ \notag
&+\frac{\epsilon^3}{6}  \int d^dx  \sqrt{\bar g(x)}  \left\{\frac{1}{8}\left(2h^{\rho\sigma}(x)h_{\rho\sigma}(x)-h^2(x) \right)h^{\mu\nu}(x)T_{\mu\nu}(x) \right. \\ \notag 
&\left.- \int d^d y \sqrt{\bar{g}(y)} \left[\frac{1}{2} h(x) h^{\mu\nu}(x)h^{\alpha\beta}(y)\frac{\delta T_{\mu\nu}(x)}{\sqrt{\bar{g}(y)}\delta g^{\alpha\beta}(y)} \right. \right. \\ &\left. \left. -\int d^dz \sqrt{\bar{g}(z)}\frac{1}{2}h^{\mu\nu}(x)h^{\rho\sigma}(y) h^{\alpha\beta}(z) \frac{\delta^2 T_{\mu\nu}}{\sqrt{\bar{g}(y)}\sqrt{\bar{g}(z)}\delta g^{\rho\sigma}(y)\delta g^{\alpha\beta}(z)}\right]\right\} +\mathcal{O}(\epsilon^4)\;,
\,
\end{align}
where, as usual with $g$ and $\bar{g}$, we denote the determinant of the respective metrics, and we have defined $h^{\mu\nu}\equiv-\bar{g}^{\mu\alpha} \bar{g}^{\nu\beta}h_{\alpha \beta}$, $h\equiv\bar{g}^{\mu\nu}h_{\mu\nu}$.\footnote{Note that with this notation, the inverse metric reads: $g^{\mu\nu}=\bar{g}^{\mu\nu}+  \epsilon\, h^{\mu\nu}+\mathcal{O}(\epsilon^2)$.} We have also used the definition of the stress-energy tensor
\begin{align}\label{Tmunudef}
T_{\mu\nu}\equiv-\frac{2}{\sqrt{g}}\frac{\delta I[\varphi,g_{\mu\nu}] }{\delta g^ {\mu\nu}}\, .
\end{align}
Hence, the partition function on the deformed background reads
\begin{align}
Z_{\epsilon}=\int \mathcal{D} \varphi\, e^{-I[\varphi,\bar{g}_{\mu\nu}]+\frac{\epsilon}{2}\int_{\mathcal{M}} d^dx \sqrt{\bar{g} }\, h^{\mu\nu}(x)T_{\mu\nu}(x)+\mathcal{O}(\epsilon^2)} \, .
\end{align}
Now, let us further assume that the free energy is an analytic function of $\epsilon$, \ie that we can expand it as
\begin{equation}\label{conj2}
\mathcal{F}(\epsilon)=\sum_{n=0}^{\infty} \frac{\epsilon^n}{n!} \mathcal{F}^ {(n)}(0)\, ,
\end{equation}
where $\mathcal{F}^ {(n)}(0) \equiv \frac{d^n\mathcal{F}}{d\epsilon^n}|_{\epsilon=0}$. A straightforward calculation of the leading correction to $\mathcal{F}(0)$ yields
\begin{equation}\label{conj33}
\mathcal{F}^{\prime}(0)=-\frac{1}{2}\frac{\int \mathcal{D} \varphi\,\left[\int_{\mathcal{M}} d^dx \sqrt{\bar{g} }\, h^{\mu\nu}(x)T_{\mu\nu}(x)\right] e^{-I[\varphi,\bar{g}_{\mu\nu}]}}{\int \mathcal{D} \varphi\, e^{-I[\varphi,\bar{g}_{\mu\nu}]}}=-\frac{1}{2}\,\int_{\mathcal{M}}  d^dx \sqrt{\bar{g} }\,h^{\mu\nu}(x)\braket{T_{\mu\nu}(x)}_{\mathcal{M}}   \, .
\end{equation}
Whenever the one point function of the energy momentum tensor in the CFT on the undeformed space vanishes
\begin{equation}\label{t0}
\braket{T_{\mu\nu}(x)}_{\mathcal{M}}=0\, , 
\end{equation}
one has $\mathcal{F}^{\prime}(0)=0$, and thus the background associated to $\bar{g}_{\mu\nu}$ is a local extremum of the free energy. From now on we will assume that \eqref{t0} holds. This is the case for odd-dimensional CFTs placed on a background metric $\bar{g}_{\mu\nu}$ on $\mathcal{M}$ which is conformally flat. An example of this situation which will be of particular interest in our work is the round sphere in odd dimensions. The fact that $\mathcal{F}^{\prime}(0)=0$ for small perturbations around the round sphere is in the spirit of the $F$-theorem in three dimensions \cite{Jafferis:2011zi,Klebanov:2011gs}. Note, however, that the two statements are not identical since we are considering deformations of the CFT with the energy momentum tensor, which is not a scalar operator.

Assuming that \eqref{t0} holds, one finds the following expression for the coefficient of the second order correction to the free energy
\begin{align}\label{eq:Fdoublep}
\mathcal{F}''(0)=&-\frac{1}{4}\int d^d y \sqrt{\bar g}\int d^d x \sqrt{\bar g}\,h^{\mu\nu}(x)h^{\rho\sigma}(y)  \left[  \braket{T_{\mu\nu}(x)T_{\rho\sigma}(y)}_{\mathcal{M}}+\underbrace{ 2  \langle\frac{\delta T_{\mu\nu}(x)}{\sqrt{\bar{g}(y)}\delta g^{\rho\sigma}(y)}\rangle_{\mathcal{M}}}\right]
\, .
\end{align}
Analogously, the coefficient of the third order correction reads
\begin{align}\label{eq:Ftripplep}
\mathcal{F}'''(0)=&-\int d^d z \sqrt{\bar g}\int d^d y \sqrt{\bar g}\int d^d x \sqrt{\bar g}\,h^{\mu\nu}(x)h^{\rho\sigma}(y)h^{\alpha\beta}(y)  \left[\frac{1}{8}\braket{T_{\mu\nu}(x)T_{\rho\sigma}(y)T_{\alpha\beta}(z)}_{\mathcal{M}}\right.\\ \notag
&+\left. \underbrace{\frac{1}{2} \langle \frac{\delta^2 T_{\mu\nu}}{\sqrt{\bar{g}(y)}\sqrt{\bar{g}(z)}\delta g^{\rho\sigma}(y)\delta g^{\alpha\beta}(z)}\rangle_{\mathcal{M}}} +\frac{3}{4} \langle\frac{\delta T_{\mu\nu}(x)}{\sqrt{\bar{g}(y)}\delta g^{\rho \sigma}(y)}T_{\alpha\beta}(z)\rangle_{\mathcal{M}}\right]\\ \notag & -\frac{1}{2}\int d^d y\sqrt{\bar g}\int d^d x \sqrt{\bar g} \, h(x) h^{\mu\nu}(x) h^{\rho\sigma}(y) \left[ \frac{3}{4}\braket{T_{\mu\nu}(x)T_{\rho\sigma}(y)}_{\mathcal{M}}+\underbrace{\langle\frac{\delta T_{\mu\nu}(x)}{\sqrt{\bar{g}(y)}\delta g^{\rho\sigma}(y)}\rangle_{\mathcal{M}}}\right]
\, .
\end{align}
If the CFT at hand lives in an odd number of dimensions, and the metric $\bar{g}_{\mu\nu}$ on $\mathcal{M}$ is conformally flat, one can show that the terms underlined with a bracket in \eqref{eq:Fdoublep} and \eqref{eq:Ftripplep} vanish. This is due to the fact that one point functions of local operators in CFTs should vanish in the absence of conformal anomalies.

Since we are interested in deformations of odd-dimensional round spheres, from now on we will ignore the underlined terms in \eqref{eq:Fdoublep} and \eqref{eq:Ftripplep}. In that case, the leading correction to the free energy is quadratic in $\epsilon$ with coefficient given by 
\begin{equation}\label{conj12}
\mathcal{F}^{\prime\prime}(0)=-\frac{1}{4} \int_{\mathcal{M}} \sqrt{\bar{g} }\, d^dx \int_{\mathcal{M}} \sqrt{\bar{g} }\, d^dy \,\left[ h^{\mu\nu}(x)h^{\rho\sigma}(y)\braket{T_{\mu\nu}(x)T_{\rho\sigma}(y)}_{\mathcal{M}} \right]  \, .
\end{equation}
Observe that if $\bar{g}_{\mu\nu}$ is the flat metric on $\mathbb{R}^d$, \req{t0} is automatically satisfied and $\ctt$ controls the leading correction to $\mathcal{F}(0)$, as it is obvious from \req{t2p}. Now let us take $\mathcal{M}$ to be a (locally) conformally flat manifold, \ie let us assume it to be related to $\mathbb{R}^d$ through a conformal mapping 
\begin{equation}\label{conj123}
f:\mathcal{M}\longrightarrow \mathbb{R}^d\, .
\end{equation}
Then, the metrics of both spaces are related through
\begin{equation}\label{conj121}
f^* (ds^2_{\mathbb{R}^d})=\Omega^2(x)ds^2_{\mathcal{M}}\, ,
\end{equation}
where $f^*$ is the pullback and $\Omega^2(x)$ the corresponding conformal factor. In odd-dimensional theories, the stress tensors for the theories on $\mathcal{M}$ and $\mathbb{R}^d$ are related through\footnote{In even dimensions, \req{tens} generically receives an anomalous contribution --- see \eg \cite{Perlmutter:2013gua}.}
\begin{equation}\label{tens}
T_{\mu\nu}(x)=\Omega^{d-2}\,M^{ab}_{\mu\nu}\,T_{ab}(X) \, ,\quad \text{where} \quad M^{ab}_{\mu\nu}\equiv \frac{\partial X^{a}}{\partial x^{\mu}}\frac{\partial X^{b}}{\partial x^{\nu}}\,,
\end{equation}
and where we denoted the coordinates in $\mathcal{M}$ and $\mathbb{R}^d$ by $\{x^{\mu}\}$ and $\{X^{a}\}$ respectively. For the correlator in the integrand of \req{conj12}, one finds
\begin{equation}\label{tens1}
\braket{T_{\mu\nu}(x)T_{\rho\sigma}(y)}_{\mathcal{M}}=\Omega^{d-2}(x)\Omega^{d-2}(y)\,M^{ab}_{\mu\nu}(x)M^{cd}_{\rho\sigma}(y)\,\braket{T_{ab}(X)T_{cd}(Y)}_{\mathbb{R}^d} \, .
\end{equation}
This expression allows us to write the two-point function on $\mathcal{M}$ in terms of the two-point function on $\mathbb{R}^d$, whose explicit expression is given by \req{t2p}. 

To recap, we find that on any odd-dimensional manifold $\mathcal{M}$ for which the map $f$ in \req{conj123} exists, the free energy behaves, under a small deformation of the metric, as
\begin{equation}\label{conj2}
\mathcal{F}(\epsilon)=\mathcal{F}(0)+\frac{\epsilon^2}{2}\mathcal{F}^{\prime\prime}(0)+\mathcal{O}(\epsilon^3)\, ,
\end{equation}
where
\begin{equation}\label{conj2t}
\mathcal{F}^{\prime\prime}(0)=\frac{-\ctt}{4\mathcal{S}^2_{d-1}} \int_{\mathcal{M}} \sqrt{\bar{g} }\, d^dx \int_{\mathcal{M}} \sqrt{\bar{g} }\, d^dy \,\left[ h^{\mu\nu}(x)h^{\rho\sigma}(y)\Omega^{d-2}(x)\Omega^{d-2}(y)\,M^{ab}_{\mu\nu}(x)M^{cd}_{\rho\sigma}(y)\,\frac{\mathcal{I}_{ab,cd}(X-Y)}{|X-Y|^{2d}}\right]  \, ,
\end{equation}
and where in the last term it is understood that we need to write all contributions as functions of $\{x\}$ and $\{y\}$ using the map $f$ defined in \req{conj123}.

In the rest of the paper, we will focus on three- and five-dimensional squashed spheres, which provide a class of deformed conformally flat manifolds. For these examples we will be able to perform the integrals in \req{conj2t} analytically and thus obtain a simple explicit expression for $\mathcal{F}^{\prime\prime}(0)$.\footnote{Note that the integral in \eqref{conj2t} suffers from UV singularities at coincident points $x \to y$. We use the same regularization procedure as in \cite{Cardy:1988cwa,Klebanov:2011gs} to remove such divergences. See Section 3 of  \cite{Closset:2012vg} for a more detailed discussion of this type of divergences.}

A few comments about the third order correction $\mathcal{F}'''(0)$ are in order. The terms on the first and third lines of \eqref{eq:Ftripplep} which are not underlined depend on the particular CFT of interest only through the constants $\ctt$, $t_2$, and $t_4$ appearing in the three-point function of the energy momentum tensor --- see appendix \ref{App:T3pt} for details. Thus, these contributions are in principle computable for general CFTs. However, the one on the second line depends on the details of the CFT at hand and, in particular, on OPE coefficients of local operators with the energy momentum tensor. This leads to the conclusion that it seems hard to obtain an explicit expression for $\mathcal{F}'''(0)$ valid for general CFTs. In the examples discussed below, we will be able to estimate $\mathcal{F}'''(0)$ for squashed spheres in some specific CFTs, including free theories as well as theories with a weakly curved holographic dual.

We would like to emphasize that in our discussion we have treated the free energy as a real quantity. In Euclidean signature this is imprecise and in general one should worry about possible contact terms with imaginary coefficients in the energy-momentum tensor correlators.\footnote{We would like to thank Yifan Wang for useful correspondence on this topic.} These type of contact terms are discussed in some detail in \cite{Closset:2012vg,Closset:2012vp} for three-dimensional CFTs on curved manifolds. In this context the relevant finite contact term for our discussion has an imaginary coefficient and is given by
\begin{equation}
\frac{i}{192\pi}\int_{\mathcal{M}}d^3x\sqrt{g}\,\varepsilon^{\mu\nu\rho} \text{tr}\left(\omega_{\mu}\partial_{\nu}\omega_{\rho}+\frac{2}{3}\omega_{\mu}\omega_{\nu}\omega_{\rho}\right)\;,
\end{equation}
where $\omega_{\mu}$ is the spin connection on $\mathcal{M}$. This gravitational Chern-Simons term can be added to the Lagrangian with an arbitrary integer coefficient and will in general affect the imaginary part of the free energy of the CFT at hand. One can also imagine that a similar gravitational Chern-Simons term built out of the spin connection can be added also to CFTs in higher odd dimensions. We expect that these counterterms will in general have an imaginary coefficient in Euclidean signature and thus they will not affect the real part of the squashed sphere free energy which is the main focus of our work.

\section{Three-dimensional squashed spheres}
\label{sec:3}

In this Section, we study the free energy of three-dimensional conformal field theories on a squashed sphere as a function of the squashing parameter. 
In this case, the metric in \req{squashedSd} becomes\footnote{Note that in the introduction and in appendix \ref{TNn} we write $ds_{S^{3}_{\alpha}}^2$ in a different set of coordinates which is more suitable for the systematic treatment of the higher-dimensional $\mathbb{CP}^{\frac{d-1}{2}}$ metrics --- see \req{cp11}. The coordinate change is given by the simple relations $\theta=2\xi_1$, $\phi=\psi_1$, $\tilde{\psi}= -\psi_1-2\psi $.} 
\begin{align}\label{sq3}
ds_{S^{3}_{\alpha}}^2& =\frac{1}{4}\left[ ds^2_{\mathbb{CP}^{1}}+\frac{1}{1+\alpha}(d\tilde{\psi}+\cos \theta d\phi)^2\right]\, , 
\end{align}
where $ds^2_{\mathbb{CP}^1}$ is the usual round metric on $S^2$,
\begin{align}\label{sq43}
ds^2_{\mathbb{CP}^{1}}=d\theta^2+\sin^2\theta d\phi^2\, ,
\end{align}
and the coordinate range is $0\leq \theta\leq \pi$, $0\leq\phi\leq 2\pi$, $0\leq \tilde{\psi} \leq 4\pi$. Note that in this parametrization we have the following relation for the small parameter $\epsilon$ defined in \eqref{pepe} 
\begin{equation}
\epsilon = - \frac{\alpha}{1+\alpha}\;.
\end{equation}
At first order in $\alpha$ (which is also first order in $\epsilon$), we can write \req{sq3} as in \req{pepe}, where $\bar{g}_{\mu\nu}$ is the metric of the round sphere
\begin{align}\label{TN1n222}
ds^2 _{S^3}=\frac{1}{4}\left[d\theta^2+d\phi^2+d\tilde{\psi}^2+2\cos\theta d\phi d\tilde{\psi} \right]\,.
\end{align}
For this special choice of deformation of the round sphere, the only non-vanishing component of $h^{\mu\nu}$ reads
\begin{equation}\label{hpsipsi}
h^{\tilde{\psi}\tilde{\psi}}=-4\, .
\end{equation}
Clearly, this is a very symmetric squashing deformation of the metric on the round sphere, which will allow for the explicit results below.

To evaluate the integral in \eqref{conj2t}, we need the explicit form of transformations in \eqref{conj121} and \eqref{tens}. To find this, we employ a conformal mapping ($f^{-1}$) from $\mathbb{R}^3$ to $S^3$ as follows
\begin{align}
X^1=  \frac{\cos\left(\frac{\phi-\tilde{\psi}}{2}\right)\sin\left(\frac{\theta}{2}\right)}{1-\sin\left(\frac{\phi+\tilde{\psi}}{2}\right)\cos\left(\frac{\theta}{2}\right)}\, ,~~~   
X^2=  \frac{\sin\left(\frac{\phi-\tilde{\psi}}{2}\right)\sin\left(\frac{\theta}{2}\right)}{1-\sin\left(\frac{\phi+\tilde{\psi}}{2}\right)\cos\left(\frac{\theta}{2}\right)}\, , ~~~
X^3= \frac{\cos\left(\frac{\phi+\tilde{\psi}}{2}\right)\cos\left(\frac{\theta}{2}\right)}{1-\sin\left(\frac{\phi+\tilde{\psi}}{2}\right)\cos\left(\frac{\theta}{2}\right)}\, .
\end{align}
In these coordinates, the metric of the round $S^3$ \req{TN1n222} is related to the flat metric on $\mathbb{R}^3$ through
\begin{align}\label{3sphmet}
f_*(ds^2_{\mathbb{R}^3})=\Omega^{2}ds^2 _{S^3}\ , \quad \text{where} \quad 
\Omega=\frac{1}{2}\left[1+\delta_{ab}X^aX^b\right]\, .
\end{align}
With these ingredients, we are ready to evaluate $\mathcal{F}^{\prime\prime}_{S^{3}_{\alpha}}(0)$ using \req{conj2t}. A somewhat tedious, but otherwise straightforward, calculation yields 
\begin{equation}\labell{f45}
\mathcal{F}^{\prime\prime}_{S^{3}_{\alpha}}(0)=-\frac{\pi^2}{48}\ctt\, .
\end{equation}
Notice that for unitary CFTs, $\ctt$ is a positive number, which in turn implies that the partition function on the round $S^{3}$ is a local maximum in the space of squashing deformations considered here. Similar observations were made in the case of CFTs with an Einstein holographic dual, a free scalar theory and the interacting $O(N)$ model in three dimensions in \cite{Hartnoll:2005yc,Anninos:2012ft}. Our result shows that this holds for general CFTs and, furthermore, that the leading correction to the round sphere result, which is always quadratic in the deformation, is controlled by the central charge $\ctt$ \req{f45}.

From these results, we can conclude that for general three-dimensional CFTs, the free energy on a squashed three-sphere with the metric \eqref{sq3} is given by
\begin{equation}\labell{f48}
\mathcal{F}_{S^{3}_{\alpha}}=\mathcal{F}_{S^{3}}-\frac{\pi^2\ctt}{96}\epsilon^2+\mathcal{O}\left(\epsilon^3\right)=\mathcal{F}_{S^{3}}-\frac{\pi^2\ctt}{96}\frac{\alpha^2}{(1+\alpha)^2}+\mathcal{O}\left(\frac{\alpha^3}{(1+\alpha)^3}\right)\, ,
\end{equation}
for small values of the squashing parameter $\alpha$ (or $\epsilon$). 

In the following subsections, we study $\mathcal{F}_{S^{3}_{\alpha}}$ for general values of $\alpha$, \ie beyond the quadratic approximation, for free CFTs as well as for theories with weakly coupled gravitational duals. The results we find, in the regime of small $\alpha$, are in harmony with the general result in \req{f48}. 

\subsection{Conformally-coupled scalar}
\label{ffc}  

Let us consider a conformally-coupled scalar field $\phi$. In order to compute the free energy $\mathcal{F}_{\textrm{sc},\,  S^3_{\alpha}}$, we need to evaluate the partition function 
\begin{align}\label{Zscalar3d}
Z_{\rm sc}=\int \mathcal{D} \phi\, e^{-\frac{1}{2}\int d^3x \sqrt{g} \left[ (\partial \phi)^2 +\frac{ R\phi^2}{8} \right]} \, ,
\end{align}
on the squashed-sphere background \req{sq3}.\footnote{The factor of $1/8$ in front of the Ricci scalar in \eqref{Zscalar3d} is the usual conformal coupling in three dimensions. In general dimension this prefactor is $\frac{d-2}{4(d-1)}$.} After a Gaussian integration, we find
\begin{align}
\mathcal{F}_{\rm{sc}}=- \log Z_{\rm{sc}}=\frac{1}{2}\log \textrm{det}\left[ \frac{-\nabla^2 +\frac{R}{8}}{ \Lambda^2}\right]\, , \label{eqn:LogZScalar3d}
\end{align}
where we have introduced the UV cutoff $\Lambda$. In order to evaluate \eqref{eqn:LogZScalar3d} on the squashed-sphere \req{sq3}, we need the eigenvalues of the conformal Laplacian on this manifold. These were used in a similar context in \cite{Dowker:1998pi,Anninos:2012ft}, and are given by\footnote{For more details, see appendix \ref{ssgd}.}
\begin{align}
\lambda_{n,q}= n^2 + \alpha (n-1-2q)^2 - \frac{1}{4(1+\alpha)} \ ,
\end{align}
with $n=1,2,\ldots $ and $q=0,1,\ldots,n-1$.

The partition function $Z_{\rm{sc}}$ has UV divergences and needs to be regularized. While it is possible to use zeta-function regularization as in \cite{Dowker:1998pi,Hartnoll:2005yc}, this leads to somewhat cumbersome calculations that are non-trivial to extend to other CFTs, like the free fermion considered in the following subsection, higher-dimensions, or to spheres with more complicated deformations. Therefore, we will use the numerical regularization methods employed in \cite{Anninos:2012ft,Anninos:2013rza,Bobev:2016sap}. The essential idea behind these is to use a ``soft regulator'', which makes it possible to split the free energy into UV- and IR-dependent parts. The IR-part consists of a convergent sum which can be straightforwardly calculated numerically, while the UV one contains all the divergences.
The only counter-terms that can be added to the action to control these divergences are
\begin{align}
I_{\textrm{div}}= A \Lambda^3\int d^3 x \sqrt{g} + B \Lambda\int d^3 x \sqrt{g} R \ , \label{eqn:divergences}
\end{align}
where $\Lambda$ is the cutoff and $A$ and $B$ are real coefficients which depend on the details of the theory at hand. 
The regularization procedure amounts to performing a numerical fit of the on-shell action plus the counterterms in \req{eqn:divergences}. From this, we can determine the values of $A$ and $B$ which allow us to remove the unwanted divergences. More details about the numerical procedure we used can be found in appendix \ref{ff} (see also \cite{Bobev:2016sap}).

\begin{figure}[ht!]
	\centering
	\includegraphics[width=0.8\textwidth]{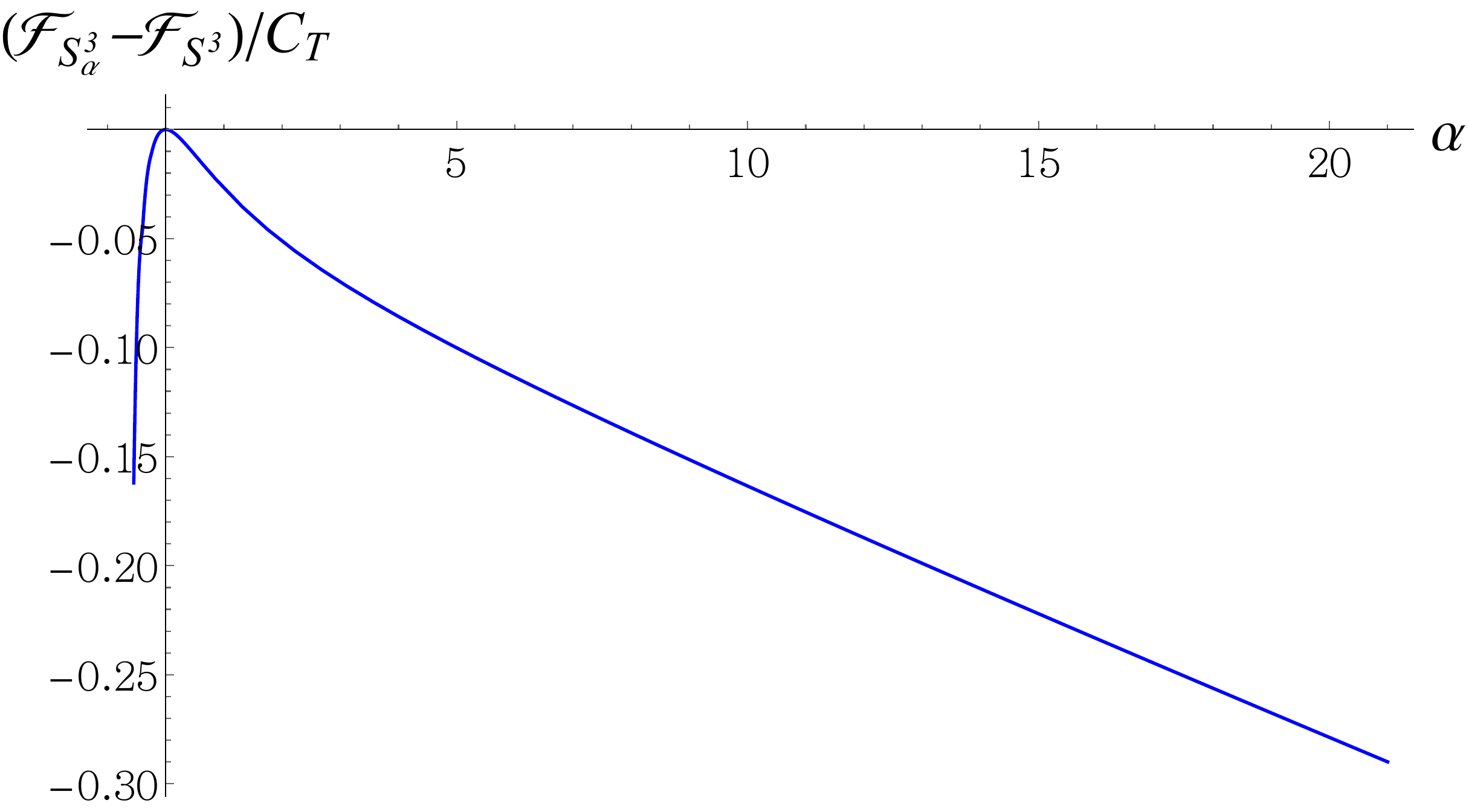}
	\caption{The free energy for a conformally coupled scalar on the squashed $S^3$ as a function of the squashing parameter $\alpha$.}\label{fig:3dscalar}
\end{figure}

After subtracting the divergences, we obtain the regularized free energy as a function of the squashing parameter $\alpha$.  We can perform this calculation for arbitrary values of $\alpha$ and our results for the free energy $\mathcal{F}_{\textrm{sc},\,  S^3_{\alpha}}$ are depicted\footnote{As explained in Section \ref{susti} we find it convenient to subtract the free energy of the round sphere and normalize the result by dividing by $\ctt$.} in Figure \ref{fig:3dscalar}. Two features of our numerical results are immediately noticeable. First, the round sphere at $\alpha=0$ leads to a local maximum of the free energy. For the large range of values of $\alpha$ that we studied, this is also a global maximum and it is natural to conjecture that this is true for all allowed values in the range $\alpha>-1$. Second, for large positive values of $\alpha$ the free energy becomes a linear function.\footnote{We are grateful to Louise Anderson and Paul McFadden for a useful correspondence on this topic.} This linear behavior for large $\alpha$ was also observed and discussed in  \cite{DeFrancia:2000xm,Hartnoll:2005yc}. In particular, in \cite{DeFrancia:2000xm}, it was argued that in the limit $\alpha \gg 1$ the free energy is given by
\begin{align}
	\mathcal{F}_{\rm{sc}}&= -\frac{\zeta(3)}{8\pi^2} \alpha -\frac{1}{24}\log(1+\alpha)+\frac{1}{12}\log(4\pi)+\zeta'(-1)-\frac{\zeta(3)}{8\pi^2}+\mathcal{O}(\alpha^{-1})\, ,\\ 
	\notag
		&\simeq-0.01522\alpha-0.04167 \log(1+\alpha)+ 0.03027\, .
\end{align}
A fit to our numerical curve yields instead
\begin{equation}
	 \mathcal{F}_{\rm{sc}}\simeq -0.01524\alpha-0.03706\log(1+\alpha)+0.06587\, .
\end{equation}
It is clear that the leading $\mathcal{O}(\alpha)$ term in our numerical results agrees very well with the result in \cite{DeFrancia:2000xm}. This is not the case, however, of the subleading contributions. It would be interesting to identify the source of these discrepancies, which do not seem attributable to numerical errors.\footnote{We note that there are some errors in the result of \cite{Dowker:1998pi} which were identified and corrected in \cite{DeFrancia:2000xm}.}

We provide further comments on the squashed sphere free energy for the free scalar and how it compares to the free energy of free fermions and holographic theories in Section \ref{susti}. However, we can readily perform some consistency checks of our numerical procedure. First, we can evaluate the free energy for $\alpha=0$. This should correspond to the well-known result for the round three-sphere free energy, given by \cite{Klebanov:2011gs} 
\begin{equation}\label{pen}
\mathcal{F}_{\textrm{sc},\,  S^3}=\frac{\log 2}{8} - \frac{3 \zeta(3)}{16 \pi^2}= 0.06380705\dots
\end{equation}
From our numerics we find $\mathcal{F}_{\textrm{sc},\,  S^3}\simeq  0.063807$, in very good agreement with \req{pen}. According to our general result \req{conj2}, the first derivative of the free energy with respect to the squashing parameter evaluated at $\alpha=0$ should vanish. Numerically, we find
$\mathcal{F}_{\textrm{sc},\,S^3_\alpha}'(0) \simeq 2.25 \cdot 10^{-11}$,
which is in clear agreement with our expectations. In addition, according to \req{f48}, we expect to find
$
-48/\pi^2\mathcal{F}_{\textrm{sc},\, S^3_\alpha}''(0)=  {\ctt}_{,\,\rm sc}=3/2.
$
Numerically, we obtain
$
-48/\pi^2 \mathcal{F}_{\textrm{sc},\,S^3_\alpha}''(0)\simeq 1.5000004
$,
which is clearly in very good agreement with the general result \eqref{f48}.

\subsection{Free fermion}
\label{sec:3dferm}

Let us now derive similar results for a free Dirac fermion.
The partition function in this case is 
\begin{align}
Z_{\rm f}=\int \mathcal{D} \psi\,  e^{-\int d^3x \sqrt{g} \left[\psi^{\dagger} (i \slashed{D} )\psi \right]} \ ,
\end{align}
where $\slashed{D}$ is the Dirac operator on the corresponding curved background. The free energy can be found by solving the Gaussian integral
\begin{align}
\mathcal{F}_{\textrm{f}} = - \log Z_{\rm f}=-\log \textrm{det}\left[ \frac{i \slashed{D}}{ \Lambda}\right] \, , \label{eqn:LogZGeneralSpinor}
\end{align}
where, once again, we introduce an energy cutoff $\Lambda$. We are interested in evaluating the determinant in \eqref{eqn:LogZGeneralSpinor} for the squashed sphere metric in \eqref{sq3}. To this end we need the eigenvalues of the Dirac operator on this background. These have been found in \cite{Dowker:1998pi, Gibbons198098, HITCHIN19741}, and read
\begin{align}\label{fermioneigen}
\lambda_{n,q,\,\pm} =   \frac{1}{\sqrt{1+\alpha}}  \pm 4\sqrt{ \frac{n^2(1+\alpha)}{4}-\alpha q (n - q) }  \, ,
\end{align}
where $n$ and $q$ are integers. For the positive branch, denoted by $+$, $n$ goes from 1 to $\infty$ and $q$ from 0 to $n$, while for the negative branch, denoted by $-$, $n$ goes from $2$ to $\infty$ and $q$ from $1$ to $n-1$. In evaluating the determinant in \eqref{eqn:LogZGeneralSpinor}, we have to sum over both branches of the eigenvalues.

\begin{figure}[ht!]
	\centering
	\includegraphics[width=0.8\textwidth]{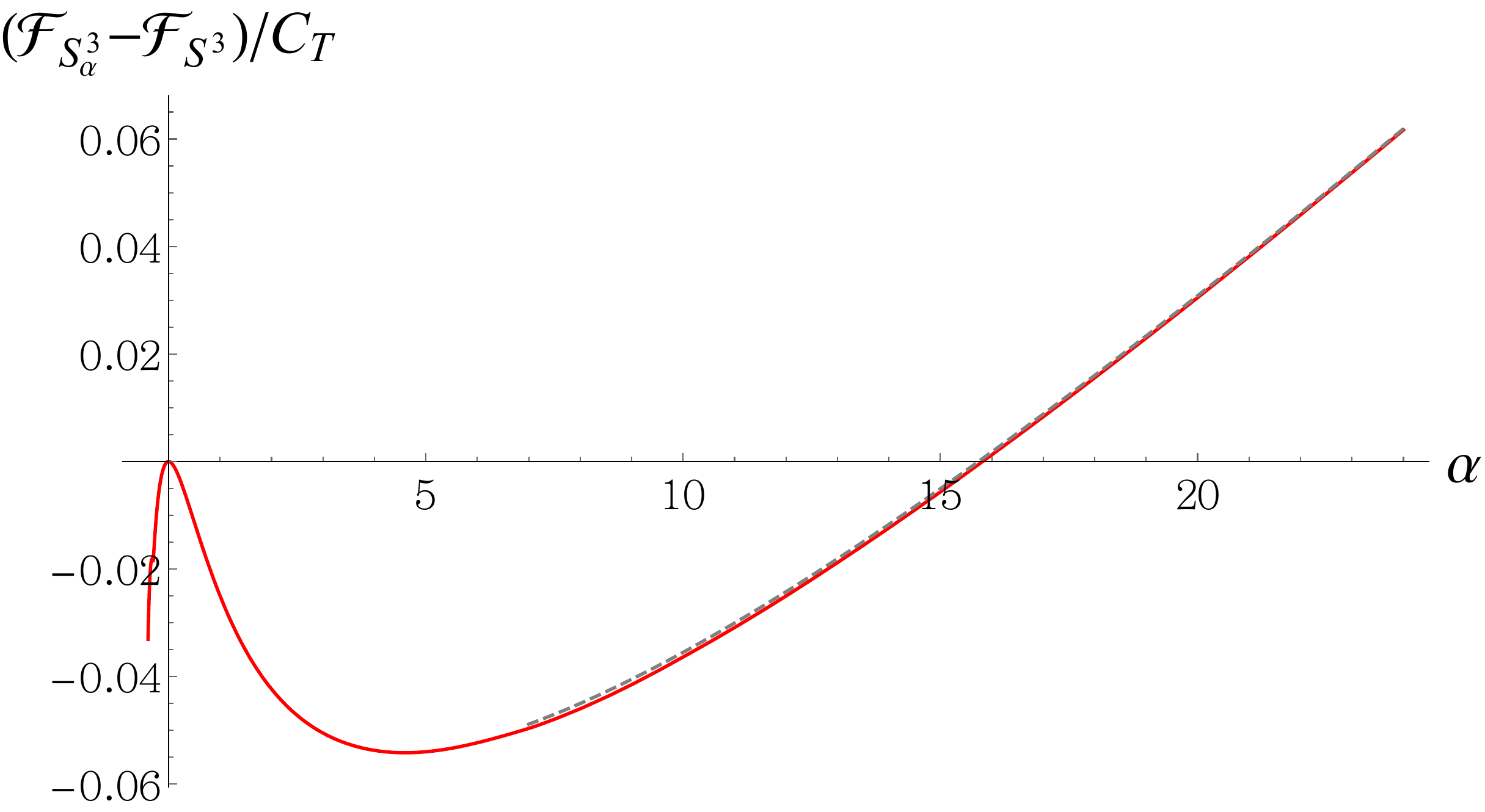}
	\caption{The free energy for a free Dirac fermion on the squashed $S^3$ as a function of the squashing parameter $\alpha$. The round sphere is not a global maximum, as the free energy grows unbounded after reaching a local minimum for $\alpha_{\rm \ssc min}\simeq  4.6228$. For large enough $\alpha$, the free energy is given by Equation \req{dow}  \cite{DeFrancia:2000xm}, which we also plot (gray dashed line). }\label{fig:3dfermion}
\end{figure}

To compute the free energy as a function of $\alpha$, we apply the same numerical procedure used in Section \ref{ffc} (and described in Appendix \ref{ff}) for the free scalar. Our numerical results are presented in Figure \ref{fig:3dfermion}. We again observe that the round sphere partition function at $\alpha=0$ is a local maximum. However, it is clearly not a global maximum since, interestingly, for large positive values of $\alpha$ we find that the free energy  increases linearly. Including subleading terms, we find 
\begin{equation}\label{line}
\mathcal{F}_{\textrm{f},\,S^3_\alpha}\simeq 0.03043 \alpha-0.16551\log(1+\alpha)+0.20929-\frac{0.07616}{(1+\alpha)}\, , \quad (\alpha \gg 1)\, .
\end{equation}
Just like for the scalar, an analytic expression was found in \cite{DeFrancia:2000xm} for $\mathcal{F}_{\textrm{f},\,S^3_\alpha}$ in this regime. This reads
\begin{align}\label{dow}
\mathcal{F}_{\textrm{f},\,S^3_\alpha}&= \frac{\zeta(3)}{4\pi^2} \alpha -\frac{1}{6}\log(1+\alpha)+\frac{1}{3}\log(4\pi)+4\zeta'(-1)+\frac{\zeta(3)}{4\pi^2}+\left(\frac{1}{96\pi^2} -\frac{1}{720}\right)\frac{16\pi^2}{(1+\alpha)}\, ,\\ 
\notag
&\simeq 0.03045 \alpha-0.16667\log(1+\alpha)+0.21244-\frac{0.05265}{(1+\alpha)}\, ,
\end{align}
which, in this case, agrees very well with our numerical result\footnote{The $\mathcal{O}(1/(1+\alpha))$ term does not agree so well, but this could be an artefact of the numerics.  } --- see also Figure \ref{fig:3dfermion}.
In contrast to the scalar case, the fermion free energy reaches a (local) minimum at an intermediate value of $\alpha$, given by
\begin{equation}\label{alphamin}
\alpha_{\rm \ssc min}\simeq 4.6228\, .
\end{equation}
%

To gain further confidence in our numerical results, we can perform some consistency checks. For $\alpha=0$, the partition function for a free Dirac fermion was evaluated analytically in \cite{Klebanov:2011gs} and reads
\begin{equation}\label{femen}
\mathcal{F}_{\textrm{f},\,S^3} = \frac{\log{2}}{4}  + \frac{3 \zeta(3)}{8 \pi^2}=0.21895948\dots
\end{equation}
With our numerical method for $\alpha=0$, we find $\mathcal{F}_{\textrm{f},\,S^3}\simeq 0.21895949$, which is in excellent agreement with \eqref{femen}. The first and second derivatives of the free energy with respect to $\alpha$, evaluated at $\alpha=0$ are given by 
$
\mathcal{F}_{\textrm{f},\,S^3_\alpha}'(0) = 9.01 \cdot 10^{-7}$ and $   -48/\pi^2 \mathcal{F}_{\textrm{f},\,S^3_\alpha}''(0)=  2.99999 ,
$ respectively. After taking into account that for the free Dirac fermion we have  ${\ctt}_{, \,\rm f}=3$, these values are in very good agreement with our general results \eqref{conj2} and \eqref{f48}. 

\subsection{Holography}
\label{holo}

Let us now study the squashed sphere free energy of three-dimensional CFTs with an Einstein gravity holographic dual using the standard tools of AdS/CFT. In order to do so, let us consider an Euclidean bulk theory of the form
\begin{equation}\label{eg}
I=-\frac{1}{16\pi G} \int_{\mathcal{M}} d^4x\, \sqrt{g} \left[\frac{6}{\ell^2}+R \right]\;,
\end{equation}
where $G$ is the Newton constant and the parameter $\ell$ sets the scale of the Anti-de Sitter (AdS) vacuum solution of the above action.

The Euclidean $AdS_4$ solution of \eqref{eg} is dual to the vacuum state of some three-dimensional CFT.\footnote{We will be agnostic about the nature of this CFT. For concreteness, one can imagine that we have obtained the action in \eqref{eg} as a consistent truncation of eleven-dimensional supergravity on $S^7$. The three-dimensional CFT is then the ABJM theory \cite{Aharony:2008ug}.} The gravitational dual of this CFT placed on the squashed sphere with metric \eqref{sq3} should be a gravitational background which is asymptotically locally (Euclidean) $AdS_4$ with a boundary metric given by \eqref{sq3}. There is indeed a class of analytic solutions to the equations of motion derived from \eqref{eg} which posses the required asymptotic structure. These are the so-called AdS-Taub-NUT and AdS-Taub-Bolt metrics which were studied in a holographic context in \cite{Hawking:1998ct,Chamblin:1998pz} (see also \cite{Bobev:2016sap} for a recent account and more complete list of references). Thus, we can apply standard holographic tools to these analytic solutions to extract information about the dual CFT theory on $S_{\alpha}^3$. 

The AdS-Taub-NUT solution is the one which is thermodynamically stable for values of $\alpha$ less than $\alpha_{\rm HP} =6 + 2\sqrt{10}\simeq 12.3246$, and thus dominates the gravitational partition function in that range. At $\alpha_{\rm HP}$, there is a Hawking-Page phase transition and the thermodynamically favored solution is AdS-Taub-Bolt, which has a non-trivial two-cycle \cite{Emparan:1999pm,Mann:1999pc}. In addition, one should note that the AdS-Taub-Bolt solutions do not exist for $\alpha<\alpha_{\rm crit}\equiv 5+3\sqrt{3}\approx 10.1962$ (see \cite{Bobev:2016sap} for a more detailed discussion). Our interest here is on relatively small values of $\alpha$, since we would like to study deformations that smoothly connect to the $AdS_4$ vacuum solution. Therefore, from now on we will focus on the AdS-Taub-NUT solution, which is reviewed in Appendix \ref{TNn}. As we show there, the free energy of this background as a function of $\alpha$ can be computed via holographic renormalization and reads \cite{Emparan:1999pm}
\begin{equation}\label{tnf1}
\mathcal{F}_{\textrm{E},\,S^3_{\alpha}}=\frac{\pi \ell^2}{2G}\frac{(1+2\alpha)}{(1+\alpha)^2}\;,
\end{equation}
where the label ``E'' stands for Einstein gravity. 

To make contact with the field theory around \eqref{f48}, we need to express this result in terms of the ``central charge'' $\ctt$. For CFTs dual to Einstein gravity with the action \eqref{eg}, one has (see \eg \cite{Buchel:2009sk} and Footnote \ref{foo})
\begin{equation}
\ctt=\frac{48\ell^2}{\pi G}\;.
\end{equation}
Using this and \req{tnf1}, it is straightforward to show that the general field theory results \req{conj2} and \req{f48} are obeyed. This is a non-trivial quantitative check of the holographic interpretation of the AdS-Taub-NUT solution. The free energy in \eqref{tnf1} takes a more suggestive form when rewritten in terms of the parameter $\epsilon$
\begin{equation}\label{sexy}
\mathcal{F}_{\textrm{E},\,S^3_{\alpha}}=\mathcal{F}_{\textrm{E},\,S^3}-\frac{\pi^2\ctt}{96}\frac{\alpha^2}{(1+\alpha)^2}=\mathcal{F}_{S^{3}}-\frac{\pi^2\ctt}{96}\epsilon^2\, .
\end{equation}
Notice that this is an exact result valid for finite\footnote{Strictly speaking, finite but smaller than the value $\alpha_{\rm HP}$ defined above \eqref{tnf1}.} values of $\alpha$ (or $\epsilon$). In particular this result implies that for three-dimensional CFTs with holographic duals captured by the action in \eqref{eg}, the quadratic approximation in \req{conj2} and \req{f48} is exact, and all higher-order corrections in powers of $\epsilon$ vanish. Notice also that $\alpha=0$ is a global maximum for the gravitational free energy \eqref{sexy}, \ie the free energy is extremized when the boundary is the round $S^3$.

\subsection{Comparison}
\label{susti}

\begin{figure}[ht!]
	\centering
	\includegraphics[width=0.8\textwidth]{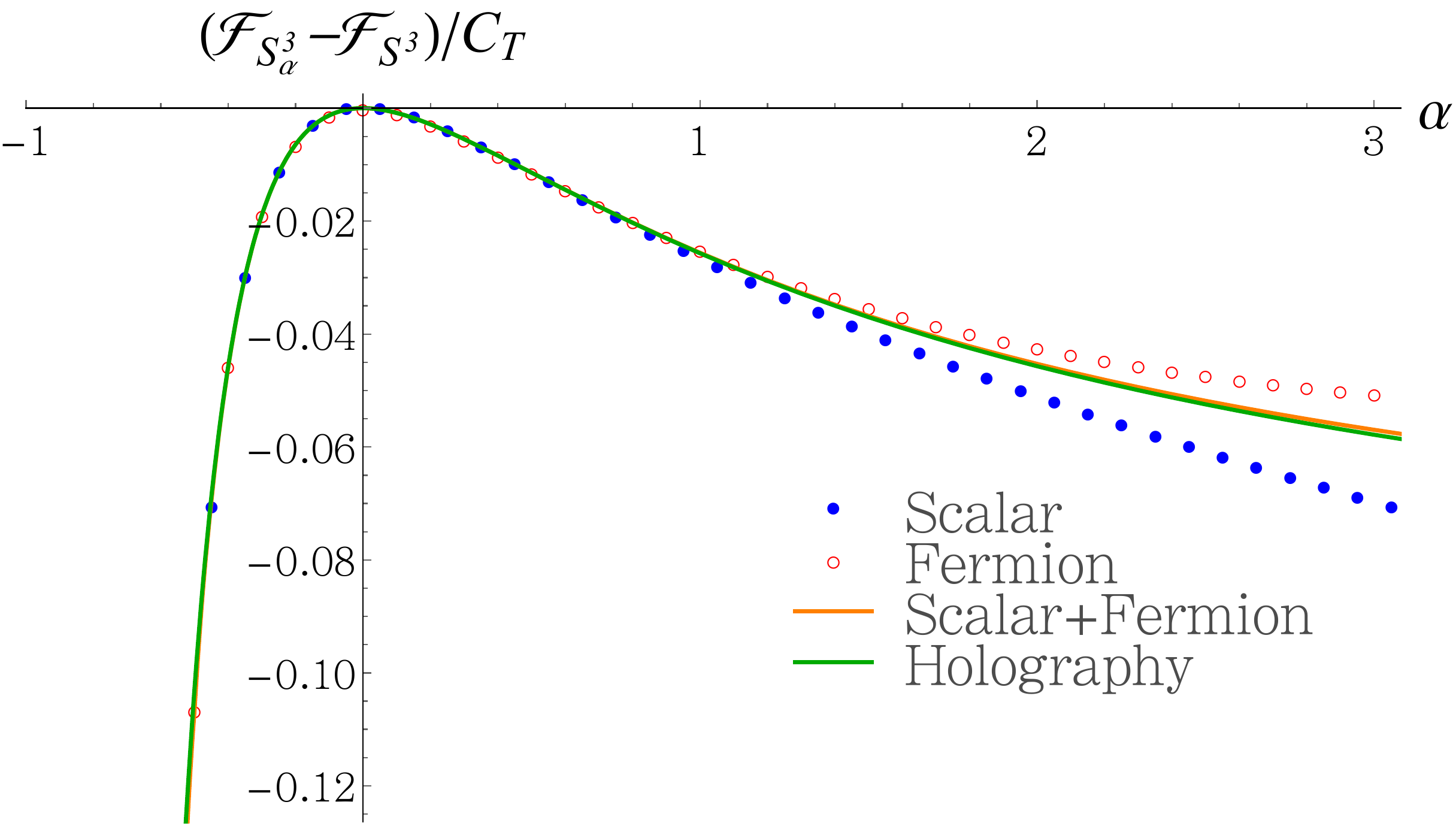}
	\caption{The normalized free energy $(\mathcal{F}_{S^3_{\alpha}}-\mathcal{F}_{S^3})/\ctt$ as a function of the squashing parameter $\alpha$ in the range $\alpha \in [-1, 3]$ for holographic theories dual to Einstein gravity (green solid line), a free scalar (blue dots), a free fermion (red dots) and the combination of both (orange solid line).}\label{fig3dalpha}
\end{figure}

\begin{figure}[h]
	\centering
	\includegraphics[width=0.8\textwidth]{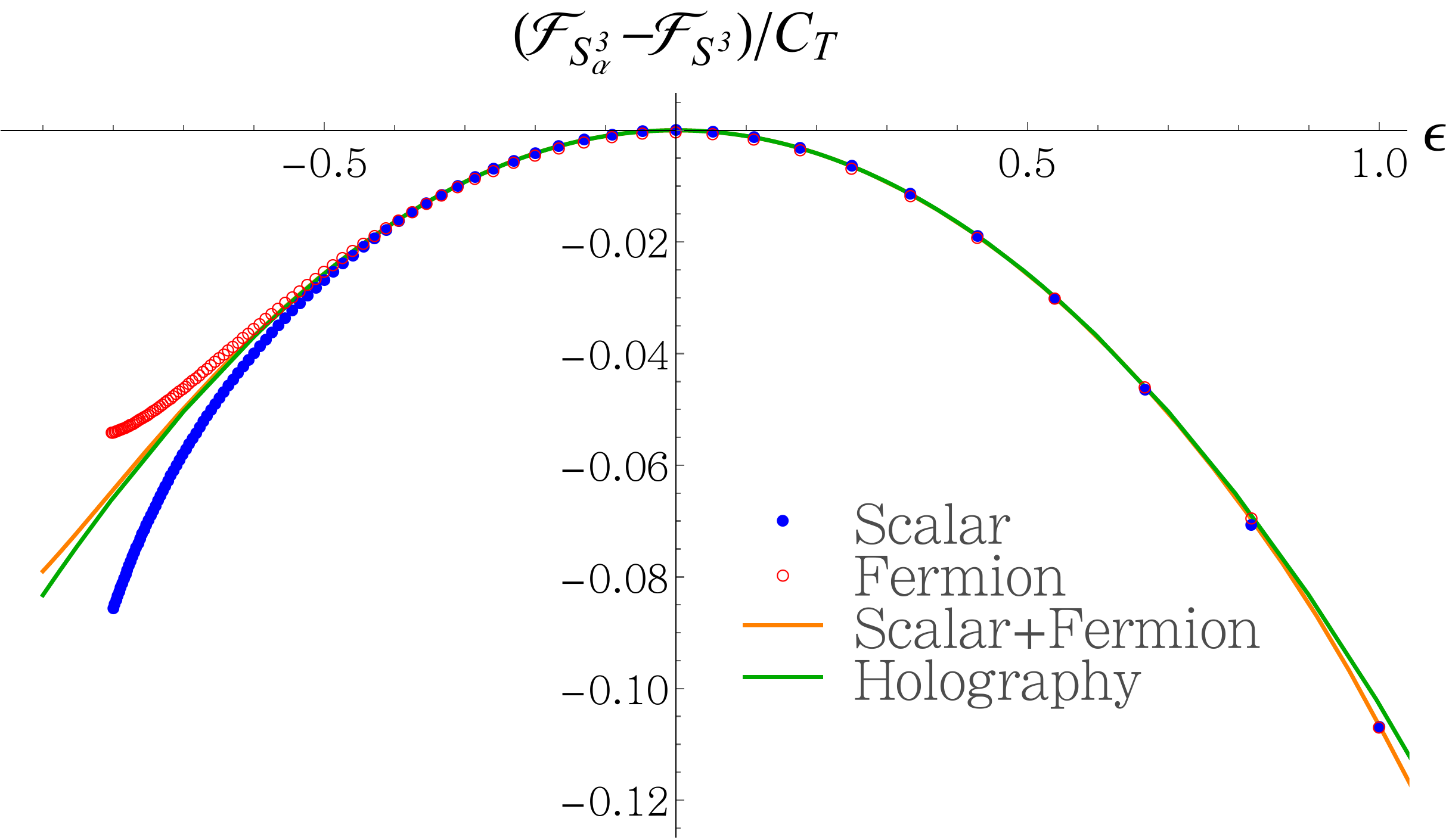}
	\caption{The normalized free energy $(\mathcal{F}_{S^3_{\alpha}}-\mathcal{F}_{S^3})/\ctt$ as a function of the squashing parameter $\epsilon$ in the range $\epsilon \in [-1, 1]$ for holographic theories dual to Einstein gravity (green solid line), a free scalar (blue dots), a free fermion (red dots) and the combination of both (orange solid line). The holographic curve is the parabola in \eqref{sexy}. }\label{fig3deps}
\end{figure}

It is time that we pause and compare the results for the free energies computed so far. We find it illuminating to study the dependence of the free energy on the squashing parameter by subtracting the value of the free energy on the round sphere and dividing by $\ctt$, \ie by considering the quantity $(\mathcal{F}_{S^3_{\alpha}}-\mathcal{F}_{S^3})/\ctt$.  Indeed, it is natural to normalize the free energy by the central charge $\ctt$, given that, as shown in Section \ref{party}, this quantity controls the leading term in the expansion around $\alpha=0$. Moreover, it is clear from \req{f48} that the round sphere value is a local maximum of $\mathcal{F}_{S^3_{\alpha}}$. In Figure \ref{fig3dalpha}, we plot this normalized free energy as a function of $\alpha$ for the three theories considered in the previous subsections. In addition to that, we present the sum of the free energy of one free boson and one free Dirac fermion. It is unclear to us why the result for this combination is so similar to the one for general CFTs with a holographic dual. The holographic result for the free energy in \eqref{sexy} is a simple parabola when parametrized in terms of the squashing parameter $\epsilon$. To this end in Figure \ref{fig3deps} we plot the free energies for the theories discussed above as a function of $\epsilon$.


Some further comments are in order. The quadratic holographic result in \eqref{sexy} is a very good approximation for the free boson and fermion in the range $\epsilon \in (-0.5,0.5)$. This is better than the expected resemblance controlled by the leading order quadratic result. To quantify this better, we can study the subleading cubic correction to $\mathcal{F}_{S^3_{\alpha}}$
\begin{equation}\label{thirde}
\frac{\mathcal{F}_{S^3_{\alpha}}-\mathcal{F}_{S^3}}{\ctt}= -\frac{\pi^2}{96} \epsilon^2 +q\epsilon^3 + \mathcal{O}\left(\epsilon^4\right)\, ,
\end{equation}
where $q\equiv\frac{\mathcal{F}^{'''}(0)}{6\ctt}$, as can be seen from \eqref{conj2}. For the free scalar and free fermion, we can find the coefficient $q$ from our numerical results. A conservative estimate leads to $|q|<10^{-3}$. Clearly, we have $q=0$ for theories with holographic duals. This suggests that the cubic correction in \eqref{thirde} is either vanishing or has a very small coefficient for general CFTs. Notice that it seems hard to calculate the number $q$ for general CFTs, since this coefficient is determined by the integral in equation \eqref{eq:Ftripplep}. The various correlation functions (in particular the two-point function terms) appearing in the integrand of \eqref{eq:Ftripplep} are theory-specific and, in general, non-vanishing. Thus, the only way to have $q=0$ is to either have a miraculous cancelation between the five terms in \eqref{eq:Ftripplep}, or to obtain a zero value of the integrand due to the contractions with the very special form of the perturbation $h^{\mu\nu}$ in \eqref{hpsipsi}. It would be very interesting to establish whether $q$ vanishes for general theories, or is simply some small non-zero number.

The fact that $\ctt$ provides the ``right'' normalization for the free energy --- in the sense that the leading contribution is universally controlled by this quantity, and that curves corresponding to very different (presumably general) theories collapse to remarkably close curves --- is reminiscent of the behavior found in \cite{Bueno1,Bueno2} for the dependence of the universal contribution to the entanglement entropy of three-dimensional CFTs from corner regions, customarily denoted $a(\theta)$, where $\theta$ is the corner opening angle. In this setup, the available results for $a(\theta)/\ctt$, corresponding to free fields, holographic theories and Wilson-Fisher fixed points of the $O(N)$ model (for $N=1,2,3$), were all shown to give rise to extremely similar curves for general values of $\theta$. Analogously to our findings for the free energy of the nearly-round squashed sphere, such agreement was observed to be exact in the almost smooth corner limit.\footnote{This was subsequently proven to be true for general holographic theories in \cite{Miao2015,Bueno4} and, finally, for arbitrary CFTs in \cite{faulkner15}.} We discuss the connection between these two, seemingly unrelated, quantities further in Section \ref{EESD}.

Another notable feature of the numerical results in Figure \ref{fig3dalpha} and Figure \ref{fig3deps} is that the free fermion and free boson curves serve as an ``envelope'' for the holographic result and the combination of both free fields. In fact, one can check that for a free CFT consisting of any number of free bosons, $N_b$, and fermions, $ N_f$, the normalized squashed sphere free energy lies between the red and the blue curves in Figures \ref{fig3dalpha} and \ref{fig3deps}. We are not able to provide a rigorous explanation of this curious fact but it is natural to offer some speculations. The higher-order terms in the expansion of $\mathcal{F}_{S^3_{\alpha}}$ around $\alpha=0$ (or equivalently $\epsilon=0$) are controlled by some combination of integrated $n$-point functions of the energy-momentum tensor. Could it be that the free boson and fermion provide an upper an lower bound for such integrated correlators? A hint that this might be true is provided by the three-point function of the energy-momentum tensor.\footnote{See appendix \ref{App:T3pt} for a review on the form of such three-point functions.} This correlator in general dimension $d$ has three independent tensor structures and is entirely fixed by three independent constants, one of which is $\ctt$ and the other two are often dubbed $t_2$ and $t_4$. For kinematical reasons, in $d=3$ one finds that $t_2=0$. The constant $t_4$ is bounded above by the value for the free boson, $t_4^{\textrm{sc}}=4$, and bounded below by the free fermion value, $t_4^{\textrm{f}}=-4$, see \cite{Osborn,Erdmenger:1996yc,Hofman:2008ar,Buchel:2009sk}.\footnote{Notice that the extensive quantities are $\ctt$ and $\ctt t_4$ (as well as  $\ctt t_2$ when $d\geq4$).} For CFTs with Einstein holographic duals, one finds $t_4=0$ \cite{Hofman:2008ar,Buchel:2009sk}. Of course, the three-point function coefficients $\ctt$ and $t_4$ should not determine the full behavior of the function $\mathcal{F}_{S^3_{\alpha}}$, since higher point functions of $T_{\mu\nu}$ are not simply determined by constants, but are complicated (and in general not known) functions of the conformal invariant cross ratios. Nevertheless, it is tempting to speculate that the free boson and fermion curves in Figures \ref{fig3dalpha} and \ref{fig3deps} provide some analogue of the Hofman-Maldacena bounds \cite{Hofman:2008ar,Buchel:2009sk}, which would be valid for odd-dimensional CFTs with a broken conformal invariance induced by the squashing deformation controlled by $\alpha$.  

It is clear from the above discussion that it would be most interesting to have access to analytic expressions for $\mathcal{F}_{S^3_{\alpha}}$ for some CFTs (including free theories) beyond the holographic result in \eqref{sexy}. While we shall not pursue this goal in the current work, let us mention in passing that a suggestive (approximate) relation between the free energies of the free boson and fermion involving the holographic result can be found as follows. Let us define $\mathcal{T}_{\alpha}\equiv (\mathcal{F}_{S^{3}_{\alpha}}-\mathcal{F}_{S^{3}})/\ctt\, +\pi^2\alpha^2/( 96(1+\alpha)^2)$, which is simply obtained by subtracting the holographic result in \eqref{sexy} from the normalized free energy. Then, within the limits of our numerical calculations, the relation $\mathcal{T}_{\rm sc, \,\alpha}\simeq-\pi^2\mathcal{T}_{\rm f, \,\alpha}/6$ holds
with a precision of $\sim 99\%$ or better for all positive values of $\alpha$ up to $\alpha \sim 8$.\footnote{Note that the factor $\zeta(2)=\pi^2/6\approx 1.645$ has been obtained by inspection and is subject to the uncertainty associated with our numerical procedure. A more precise determination of the proportionality constant might lead to a less charismatic value such as $1.646$ or $1.644$. Note also that, while it is remarkable that such a relation holds for not so small values of $\alpha$, our numerics suggests that it stops being a good approximation for $\alpha \geq 9$.}

\section{Five-dimensional squashed spheres}
\label{sec:5}

Many of the results in the previous section can be readily generalized to five-dimensional squashed spheres. The discussion is very similar, so we will be relatively brief. Let us start by applying the general results of Section \ref{party}. We can construct a conformal mapping ($f^{-1}$) from $\mathbb{R}^5$ to $S^5$ as\footnote{Observe that, as opposed to the three-dimensional case studied in Section \ref{sec:3}, here we use the same coordinates as in appendix \ref{TNn}.}
\begin{align}\notag
X^1&=  \frac{\cos\left(\xi_2 \right)\cos\left(\psi \right)}{1-\sin\left(\xi_2\right)\sin\left(\xi_1\right)\sin\left(\psi+\psi_1+\psi_2\right)}\, ,\quad   
X^2=   \frac{\cos\left(\xi_2 \right)\sin\left(\psi \right)}{1-\sin\left(\xi_2\right)\sin\left(\xi_1\right)\sin\left(\psi+\psi_1+\psi_2\right)}\, , \\ \notag
X^3&=  \frac{\sin\left(\xi_2 \right)\cos\left(\xi_1 \right)\cos\left(\psi+\psi_2\right)}{1-\sin\left(\xi_2\right)\sin\left(\xi_1\right)\sin\left(\psi+\psi_1+\psi_2\right)}\, ,\quad
X^4=    \frac{\sin\left(\xi_2 \right)\cos\left(\xi_1\right)\sin\left(\psi+\psi_2\right)}{1-\sin\left(\xi_2\right)\sin\left(\xi_1\right)\sin\left(\psi+\psi_1+\psi_2\right)}\, ,\quad   \\
X^5&=    \frac{\sin\left(\xi_2 \right)\sin\left(\xi_1 \right)\cos\left(\psi+\psi_1+\psi_2\right)}{1-\sin\left(\xi_2\right)\sin\left(\xi_1\right)\sin\left(\psi+\psi_1+\psi_2\right)}\, .
\end{align}
In this coordinates, which satisfy $0\leq\xi_{1,2} \leq \pi/2$, $-2\pi \leq \psi_1 \leq 0$, $0\leq \psi \leq 2\pi$ and $0\leq \psi_2 \leq 3\pi$, the metric of the round $S^5$ is (see Appendix \ref{TNn} for details)
\begin{align}\label{sq5}
ds_{S^{5}}^2 =d\xi_2^2+d\psi^2+ \sin ^2\xi_2 &\left[d\xi_1^2+d\psi_1^2/2-d\psi_1 \cos (2 \xi_1) \left(d\psi_1/2 + d\psi_2+d\psi \right)\right. \\ \notag  &\qquad\qquad\qquad\qquad \left.+ d\psi_2 (d\psi_2+2
  d\psi)+d\psi_1 (d\psi_2+
  d\psi)\right]\, ,
\end{align}
and is related to the usual flat metric on $\mathbb{R}^5$ through
\begin{align}\label{3sphmet}
f_*(ds^2_{\mathbb{R}^5})=\Omega^{2}ds^2 _{S^5}\, \quad \text{where} \quad 
\Omega=\frac{1}{2}\left[1+\delta_{ab}X^aX^b\right]\, .
\end{align}
The only non-vanishing component of the inverse metric perturbation $h^{\mu\nu}$ reads
\begin{equation}
h^{\psi\psi}=-1\, .
\end{equation}
Our main interest is in evaluating the free energy of CFTs on this squashed sphere background. As discussed around \eqref{Ftildedef}, it is convenient in the context of the $F$-theorem to discuss the quantity $\tilde{\mathcal{F}}_{S^{5}_{\alpha}}$ instead, which for $d=5$ differs by a sign from the usual definition of the free energy. In complete analogy with the $d=3$ case studied in Section \ref{sec:3}, we can now evaluate $\tilde{\mathcal{F}}^{\prime\prime}_{S^{5}_{\alpha}}(0)$ using the general integral in \req{conj2t}. The calculation is tedious, but the final result takes the simple form
\begin{equation}\labell{f455}
\tilde{\mathcal{F}}^{\prime\prime}_{S^{5}_{\alpha}}(0)= - \frac{3\pi^2}{320}\ctt\, .
\end{equation}
This implies that, for general CFTs, the free energy on a squashed five-sphere is given by 
\begin{equation}\labell{f485}
\tilde{\mathcal{F}}_{S^{5}_{\alpha}}=\tilde{\mathcal{F}}_{S^{5}}-\frac{3\pi^2\ctt}{640}\epsilon^2+\mathcal{O}\left(\epsilon^3\right)=\tilde{\mathcal{F}}_{S^{5}}-\frac{3\pi^2\ctt}{640}\frac{\alpha^2}{(1+\alpha)^2}+\mathcal{O}\left(\frac{\alpha^3}{(1+\alpha)^3}\right)\, ,
\end{equation}
for small values of the squashing parameter. We now proceed beyond this leading order approximation and evaluate the free energy for general values of $\alpha$ in some specific CFTs.

\subsection{Conformally-coupled scalar}
\label{freee}

The partition function for a conformally coupled scalar field in a general five-dimensional background is given by
\begin{align}
Z_{\rm sc}=\int \mathcal{D} \phi\, e^{-\frac{1}{2}\int d^5x \sqrt{g} \left[ (\partial \phi)^2 +\frac{3 R\phi^2}{16} \right]} \, .
\end{align}
As in the three-dimensional case, this leads to the following expression for the free energy:
\begin{align}
 \mathcal{F}_{\rm sc}=- \log Z_{\rm sc}=\frac{1}{2}\log \textrm{det}\left[ \frac{-\nabla^2 +\frac{3R}{16}}{ \Lambda^2}\right]\ . \label{eqn:LogZScalar5d}
\end{align}
To evaluate this determinant, we need the eigenvalues of the conformal Laplacian operator as well as their multiplicities. In Appendix \ref{ssgd}, these are worked out for a general $d$-dimensional squashed sphere for odd values of $d$. The result for the eigenvalues, $\lambda_{n,q}$, and the multiplicities, $m_{n,q}$, in $d=5$ reads
\begin{align}
\lambda_{n,q}&=(n-1)(n+3)+\alpha(n-1-2q)^2+\frac{3(5+6\alpha)}{4(1+\alpha)}\, ,\label{eigensq5d}\\
m_{n,q}&=\frac{\left(n+1\right)(q+1)(n-q)}{2}\, ,\label{degesq5d}
\end{align}
where the integers $n$ and $q$ obey $n\geq1$ and $0\leq q\leq n-1$. Plugging these eigenvalues into the free energy \eqref{eqn:LogZScalar5d}, gives rise to a diverging sum, on which we apply the same numerical regularization procedure as in the three-dimensional case. In particular, we use a soft regulator to separate the divergent and convergent parts. In five dimensions, the divergent terms are
\begin{align}
I_{\textrm{div}}= A \Lambda^5\int d^5 x \sqrt{g} + B \Lambda^3\int d^5 x \sqrt{g} R + C \Lambda \int d^5 x \sqrt{g} R^2 \, . \label{eqn:divergences2}
\end{align}
Subtracting these counterterms with appropriate coefficients, we obtain the regularized free energy as a function of the squashing parameter $\alpha$. In Figure \ref{fig:F5dalpha}, we plot our results for the free energy for a broad range of values of $\alpha$. Some additional comments on this function are provided in Section \ref{compa5}. As a consistency check of our numerical results, we can once again compare with the result for the free energy on a round $S^5$, i.e at $\alpha=0$ \cite{Klebanov:2011gs}
\begin{align}
  \tilde{\mathcal{F}}_{\textrm{sc},\, S^5} = \frac{1}{2^8} \left( 2 \log 2 +\frac{2 \zeta(3)}{\pi^2}- \frac{15\zeta(5)}{\pi^4}\right) \approx 5.74 \times 10^{-3} \, ,
\end{align}
which is in excellent agreement with our numerical result. Similarly, we find that $ \tilde{\mathcal{F}}^{\prime}_{\textrm{sc},\, S^5_\alpha}(0)\sim 10^{-10}$, which is in harmony with the general result in \eqref{f485}. The five-dimensional free scalar has $\ctt=5/4$, which is again in excellent agreement with \eqref{f485} and our numerical estimate $ -\frac{320}{3\pi^2} \tilde{\mathcal{F}}^{\prime\prime}_{\textrm{sc},\, S^5_\alpha}(0) = 1.249995 \approx \ctt $.

\subsection{Gauss-Bonnet holography}
\label{hd}

Another class of CFTs for which the squashed sphere partition function can be explicitly computed is provided by conformal theories with weakly coupled holographic duals. We will again adopt a bottom-up approach and will consider five-dimensional CFTs dual to six-dimensional Gauss-Bonnet gravity in the bulk. The Euclidean action of this theory is given by
\begin{equation}\label{gb}
I_{\rm GB}= -\frac{1}{16\pi G}\int_{\mathcal{M}}d^{6}x\, \sqrt{g} \left[\frac{20}{\ell^2}+R+\frac{\ell^2 \lambda_{\rm GB}\,\mathcal{X}_4}{6} \right]\, ,
\end{equation}
where $\mathcal{X}_4=R_{\mu\nu\rho\sigma}R^{\mu\nu\rho\sigma}-4R_{\mu\nu}R^{\mu\nu}+R^2$ is the dimensionally-extended Euler density of four-manifolds. The Gauss-Bonnet term is topological in four bulk dimensions, \ie for $d=3$,  and does not modify the gravitational equations of motion for $d\leq 3$. It does, however, contribute to the equations of motion for $d>3$. Interestingly, those equations are still second order,\footnote{Lovelock theories, of which Einstein and Gauss-Bonnet gravities are particular cases, are the most general theories constructed from contractions of the metric and the Riemann tensor that posses covariantly divergence-free second-order equations of motion \cite{Lovelock2}.} which in turn allows for the construction of many non-trivial analytic solutions of this theory in various dimensions.  In particular, AdS-Taub-NUT and Bolt solutions to \req{gb} have been constructed not only for $d=5$, but for general odd values of $d$ in \cite{Awad:2000gg}  (when $\lambda_{\rm GB}=0$) and in \cite{Dehghani:2005zm} (for general values of $\lambda_{\rm GB}$). The boundary metric of the class of NUT solutions relevant for our purposes is precisely that of a squashed-sphere $S^5_{\alpha}$ \req{squashedSd} with (see Appendix \ref{TNn} for details)
\begin{equation}\label{sqgb}
\frac{n^2}{\ell^2}=\frac{1}{6f_{\infty}(1+\alpha)}\, , 
\end{equation}
where we have defined
\begin{equation}\label{finfdef}
f_{\infty}\equiv \frac{2}{(1+\sqrt{1-4\lambda_{\rm GB}})}\;,
\end{equation}
 and the Gauss-Bonnet coupling is constrained to obey $\lambda_{\rm GB}\leq 1/4$.\footnote{For $\lambda>1/4$ there is no well-defined $AdS_6$ vacuum of the theory defined in \eqref{gb} and we thus disregard this range.} The free energy of these solutions was computed in \cite{Clarkson:2002uj,KhodamMohammadi:2008fh} and can be written as\footnote{We compute this free energy in Appendix \ref{TNn}. We find an extra overall factor of $9/8$ with respect to the result in \cite{Clarkson:2002uj,KhodamMohammadi:2008fh}. The fact that \req{tnfgb0} satisfies the general relation \req{f485}, makes us confident that our result is correct.}
\begin{equation}\label{tnfgb0}
\tilde{\mathcal{F}}_{\textrm{GB},\, S^{5}_{\alpha}}= \frac{  \pi ^2 \ell^4\left[6\alpha^2 (f_{\infty}-1)+(1+3\alpha)(3
	f_{\infty}-4)\right]}{3G (\alpha+1)^3 f_{\infty}^3 }\;.
\end{equation}
The central charge $\ctt$ of the dual CFT can be computed using standard holographic tools, and reads \cite{Buchel:2009sk}
\begin{equation}\label{CTGB}
\ctt=\frac{640(2-f_{\infty})\ell^4}{3f_{\infty}^3G}\,.
\end{equation}
Using \eqref{tnfgb0} and \eqref{CTGB}, it is easy to show that the general relation in \req{f485} is obeyed. In the limit $\lambda_{\rm GB}\rightarrow 0$,  we have $f_{\infty}\rightarrow 1$, and we obtain the Einstein gravity result
\begin{equation}\label{tnf555}
\tilde{\mathcal{F}}_{\textrm{E},\,S^5_{\alpha}}=\frac{  \pi ^2 \ctt}{640}-\frac{\pi^2\ctt}{640}\frac{\alpha^2(\alpha+3)}{(1+\alpha)^3}\;.
\end{equation}
Observe that for $\lambda_{\rm GB}\neq 0$, we have two dimensionless quantities, namely $\ell^4/G$ and $\lambda_{\rm GB}$, which appear in the expression for the free energy \eqref{tnfgb0}. In particular, this means that in general we cannot write $\mathcal{F}_{\textrm{GB},\, S^{5}_{\alpha}}$ in terms of $\ctt$ alone, like in the case of Einstein gravity.  To understand better the result in \eqref{tnfgb0} in terms of field theory quantities, we can use some of the holographic results in \cite{Buchel:2009sk}. In particular, it is known that the three-point function of the energy-momentum tensor of CFTs dual to Gauss-Bonnet gravity has a special structure which leads to the vanishing of one of the three independent constant coefficients, namely, $t_4=0$. In addition to $\ctt$ in \eqref{CTGB}, the other non-vanishing coefficient which fully determines the $T_{\mu\nu}$ three-point function is
\begin{equation}\label{t2GBdef}
t_2=-\frac{40(f_{\infty}-1)}{3(f_{\infty}-2)}\;.
\end{equation}
With this at hand, it is possible to express \req{tnfgb0} in terms of the squashing parameter and the coefficients $\ctt$ and $t_2$ as
\begin{align}\label{tnfgb}
\tilde{\mathcal{F}}_{\textrm{GB},\, S^{5}_{\alpha}}&= \frac{  \pi ^2 \ctt}{640}\left[\left(1-\frac{3}{20}t_2\right)-  \frac{3\alpha^2}{(1+\alpha)^2}+\left(2+\frac{3}{20}t_2\right) \frac{\alpha^3}{(1+\alpha)^3}\right] \\ \notag &=\frac{  \pi ^2 \ctt}{640}\left[\left(1-\frac{3}{20}t_2\right)- 3\epsilon^2 -\left(2+\frac{3}{20}t_2\right) \epsilon^3\right]\;.
\end{align}
Notice that the leading order quadratic term in \eqref{f485} is supplemented only by a cubic term in $\epsilon$. 

In the context of holography, $\lambda_{\rm GB}$ cannot take arbitrary values. Indeed, in general, it is possible to find lower and upper bounds on the couplings of higher-derivative gravities by using certain physically sensible criteria on the corresponding dual theories --- see \eg \cite{Hofman:2008ar,Buchel:2009sk,Buchel:2009tt,deBoer:2009pn,Ge:2009eh,Camanho:2009vw,Banerjee:2014oaa,Brigante:2008gz}. The most stringent known bounds in the case of holographic theories dual to $(d+1)$-dimensional Gauss-Bonnet gravity come from requiring that the energy flux is positive everywhere for certain scattering processes or, alternatively, from imposing that causality is respected in the dual CFT \cite{Buchel:2009sk}. In the case of interest for us, \ie for $d=5$, the corresponding constraints read
\begin{equation}\label{constra}
 -\frac{51}{196}\leq \lambda_{\rm GB} \leq \frac{39}{256}\, .
\end{equation}
In Figure \ref{fig:F5dalpha} we have plotted $(\tilde{\mathcal{F}}_{\textrm{GB},\, S^5_\alpha}- \tilde{\mathcal{F}}_{\textrm{GB},\, S^5})/\ctt $ for all the values of $\lambda_{\rm GB}$ allowed by \req{constra}.

\subsection{Comparison}
\label{compa5}

\begin{figure}[H]
 	\centering
 	
 	\includegraphics[width=0.71\textwidth]{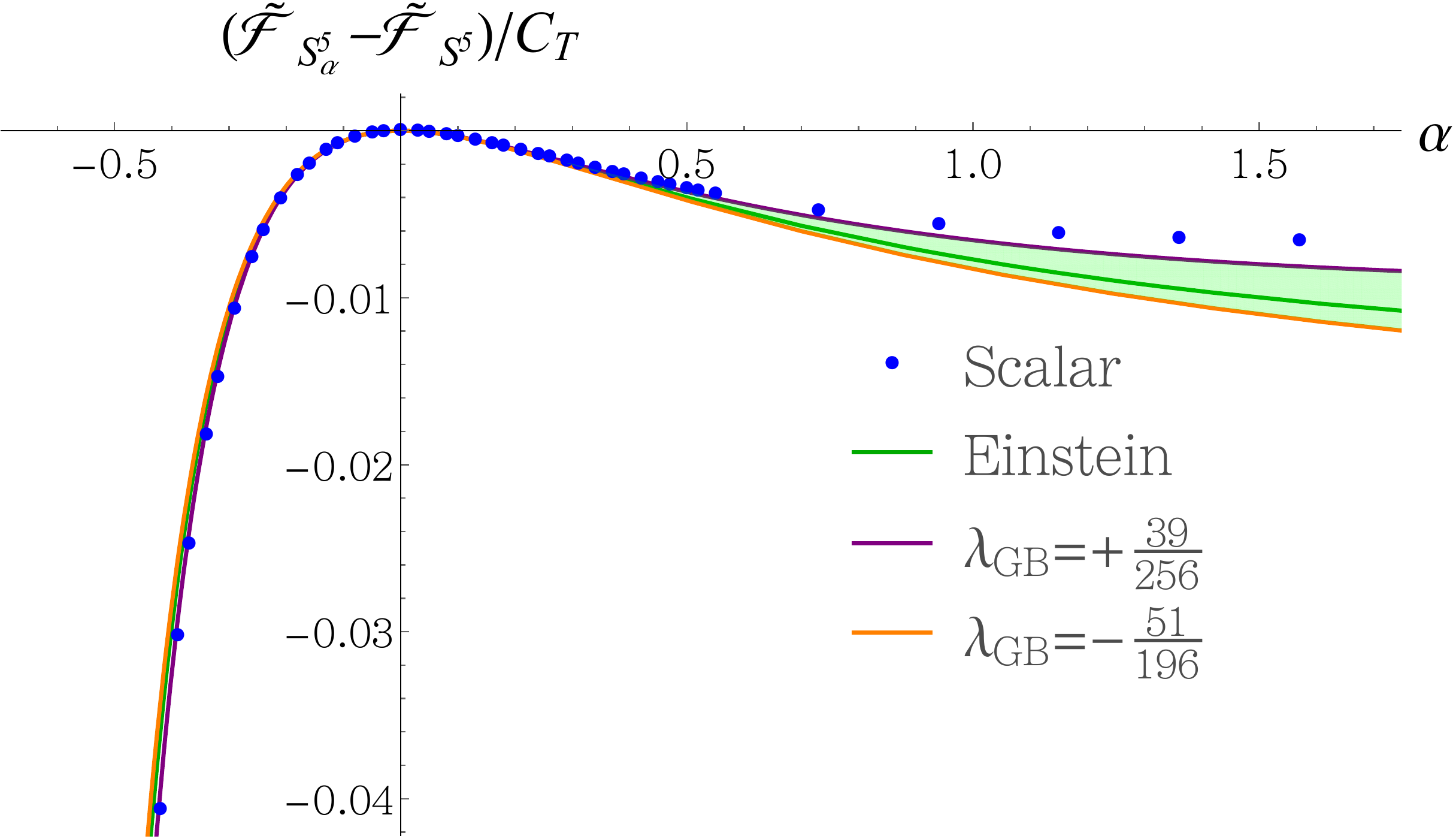}
 	
 	\caption{A plot of $(\tilde{\mathcal{F}}_{S^5_{\alpha}}-\tilde{\mathcal{F}}_{S^5})/\ctt$ as a function of $\alpha$ for a free scalar (blue dots), and for holographic theories dual to Gauss-Bonnet gravity (green shaded region). The holographic results include the Einstein theory (green) and the limiting cases \req{constra} allowed by causality in the dual CFT (purple and orange). }\label{fig:F5dalpha}
 \end{figure}

\begin{figure}[H]
	\centering
	\includegraphics[width=0.75\textwidth]{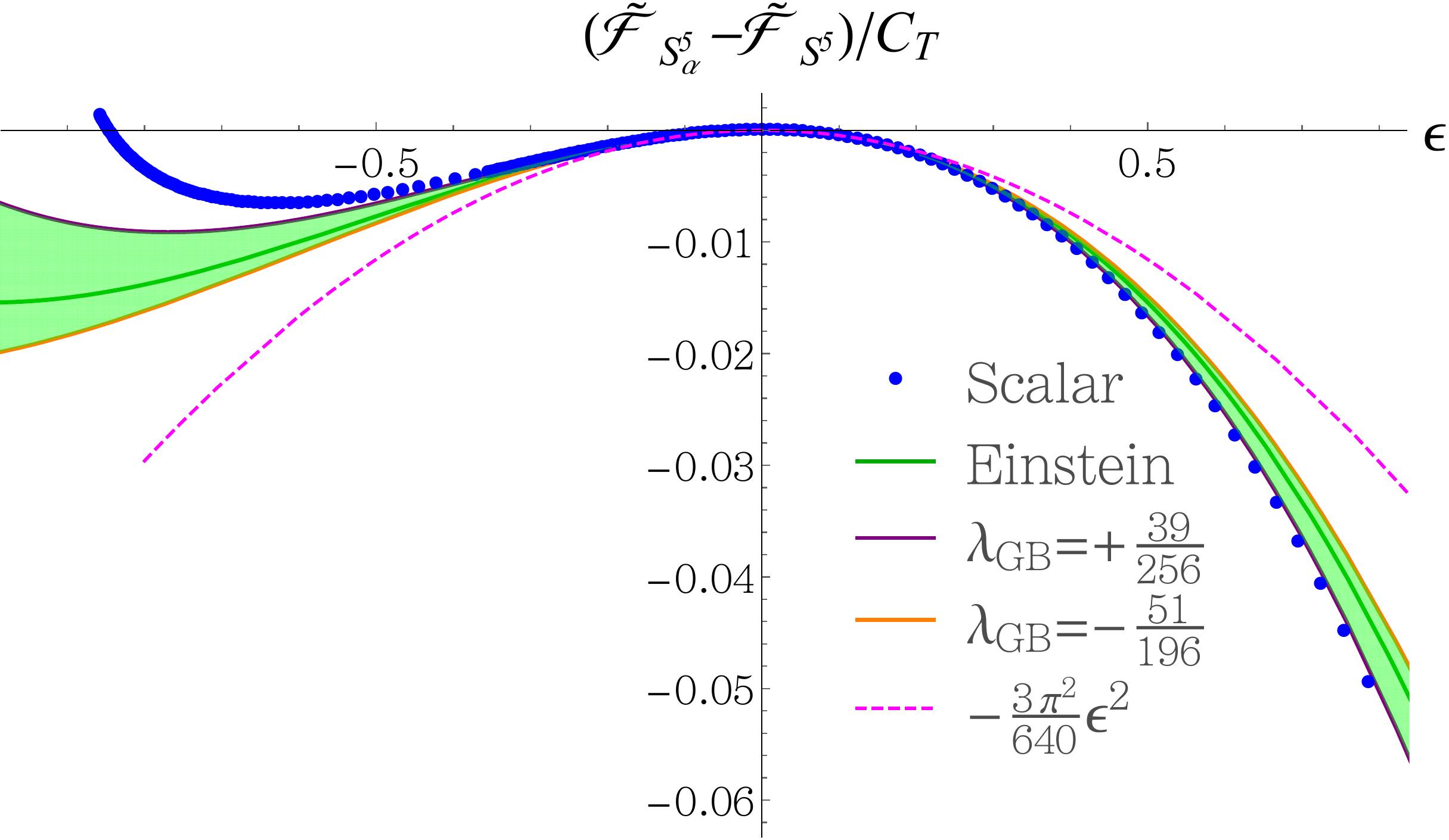}
	\caption{The same quantity as in Figure \ref{fig:F5dalpha} as a function of $\epsilon$. We have also included the leading universal quadratic term (dashed pink line).}\label{fig:F5depsilon}
\end{figure}

After computing the squashed sphere free energy for a conformally coupled scalar field and a one-parameter family of holographic theories, it is instructive to compare the results. As in Section \ref{susti}, it is useful to perform this comparison by plotting the quantity $(\tilde{\mathcal{F}}_{S^5_{\alpha}}-\tilde{\mathcal{F}}_{S^5})/\ctt$ as a function of the squashing parameter. We present such plots in Figure \ref{fig:F5dalpha} and Figure \ref{fig:F5depsilon}.

 Similarly to the three-dimensional setup, the round sphere free energy $\tilde{\mathcal{F}}_{S^5}$ is a local maximum of the function $\tilde{\mathcal{F}}_{S^5_{\alpha}}$ for all theories discussed above. This is expected, and follows from the general result in  \eqref{f485}. It is clear however from Figure \ref{fig:F5depsilon}, and from the analytic holographic result in \eqref{tnfgb}, that the round sphere free energy is not a global maximum. Another interesting feature of the curves in Figures 5 and 6 is that the analytic holographic results are very close to the numerical curve for the free boson in a surprisingly large region of values for the squashing parameter. In this case, it looks like the leading quadratic term produces an approximation which is considerably worse than its three-dimensional analog. It seems, however, that the Gauss-Bonnet result \req{tnfgb}, which contains a cubic term in $\epsilon$, does a much better job.

Some additional speculations and conjectures naturally emerge from our results in Figures \ref{fig:F5dalpha} and \ref{fig:F5depsilon}, and some of the results in \cite{Buchel:2009sk}, valid at the conformal fixed point at $\alpha=0$. In particular it was shown in \cite{Buchel:2009sk} that the following inequalities are obeyed by the coefficients $t_2$ and $t_4$, which control the three-point function of the energy-momentum tensor in unitary CFTs
\begin{equation}\label{t2t4bounds}
\begin{split}
1-\frac{t_2}{d-1}-\frac{2t_4}{d^2-1} &\geq 0\;, \\
1-\frac{t_2}{d-1}-\frac{2t_4}{d^2-1}+\frac{t_2}{2} &\geq 0\;, \\
1-\frac{t_2}{d-1}-\frac{2t_4}{d^2-1} +\frac{d-2}{d-1}(t_2+t_4)&\geq 0\;.
\end{split}
\end{equation}
For a free (conformally coupled) scalar one has $t_2^{\text{sc}} = 0$ and $t_4^{\text{sc}} = (d^2-1)/2$. For a free Dirac fermion one has (in $d>3$) $t_2^{\text{f}} = d+1$ and $t_4^{\text{f}} = -d-1$. These values saturate some of the inequalities in \eqref{t2t4bounds} and, in fact, one can ``populate'' the entire range of allowed values for $t_2$ and $t_4$ in \eqref{t2t4bounds} by taking a theory of $N_b$ free scalars and $N_f$ free fermions, and varying $N_b$ and $N_f$ appropriately. For CFTs dual to the Gauss-Bonnet theory defined by the action in \eqref{gb}, one finds that $t_4=0$, and $t_2$ is given by \eqref{t2GBdef}. This, combined with the bounds in \eqref{t2t4bounds}, leads to the bound on $\lambda_{\text{GB}}$ in \eqref{constra}. This bound on $\lambda_{\text{GB}}$ at the conformal fixed point translates into the shaded green region on the plots in Figures \ref{fig:F5dalpha} and \ref{fig:F5depsilon}. This fact naturally suggests that the unitarity bounds on $t_2$ and $t_4$ in \eqref{t2t4bounds} may extend away from the $\alpha=0$ conformal point and ultimately result in a set of ``enveloping'' curves which constrain the function $(\tilde{\mathcal{F}}_{S^5_{\alpha}}-\tilde{\mathcal{F}}_{S^5})/\ctt$ for general CFTs on a squashed sphere. Notice that the fact that the numerical results for the free boson lie outside the shaded region in Figures \ref{fig:F5dalpha} and \ref{fig:F5depsilon} is not incompatible with this conjecture, since we have $t_4=12$ for the free boson, which is greater than the $t_4=0$ value for Gauss-Bonnet. In fact, if our conjecture is true, and employing an analogy with the three-dimensional case discussed in Section \ref{susti}, it is natural to suspect that the free energy of the conformally coupled scalar is one of the putative ``enveloping'' curves. Needless to say, it will be most interesting if we can compute the function $\tilde{\mathcal{F}}_{S^5_{\alpha}}$ for any other CFTs and ultimately confirm or disprove our conjecture.

\section{Discussion} 
\label{sec:disc}   

In this section, we offer a short discussion on several topics that are related to our main results presented above. In particular, we discuss the difference between the supersymmetric free energy evaluated by supersymmetric localization and the non-supersymmetric free energy studied in this work. We also study the effect on the free energy of a generalization of the squashed sphere metric in \eqref{sq3} which preserves less isometry and depends on one more squashing parameter. Finally we comment on the resemblance between our results and recent studies of entanglement entropy for deformed entangling regions. We conclude by pointing out several interesting avenues for further study.

\subsection{Three-dimensional $\mathcal{N}=2$ SCFTs on the squashed $S^3$}

As mentioned in Section \ref{sec:intro}, supersymmetric localization is a powerful tool to extract results for supersymmetric QFT observables on curved manifolds. In particular, it is possible to compute the path integral of three-dimensional $\mathcal{N}=2$ SCFTs on the squashed $S^3$ with the metric \eqref{sq3} with this method --- see \cite{Hama:2011ea,Imamura:2011wg}. In addition to the non-trivial background metric, in this setup it is required to turn on a background field for the gauge field that couples to the superconformal $U(1)$ R-current present in every such theory. This modifies the path integral and makes the resulting supersymmetric free energy, which we will denote by $\mathfrak{F}$, not equivalent to the intrinsically non-supersymmetric free energy $\mathcal{F}$ discussed throughout this work.

\begin{figure}[ht!]
	\centering
	\includegraphics[width=0.7\textwidth]{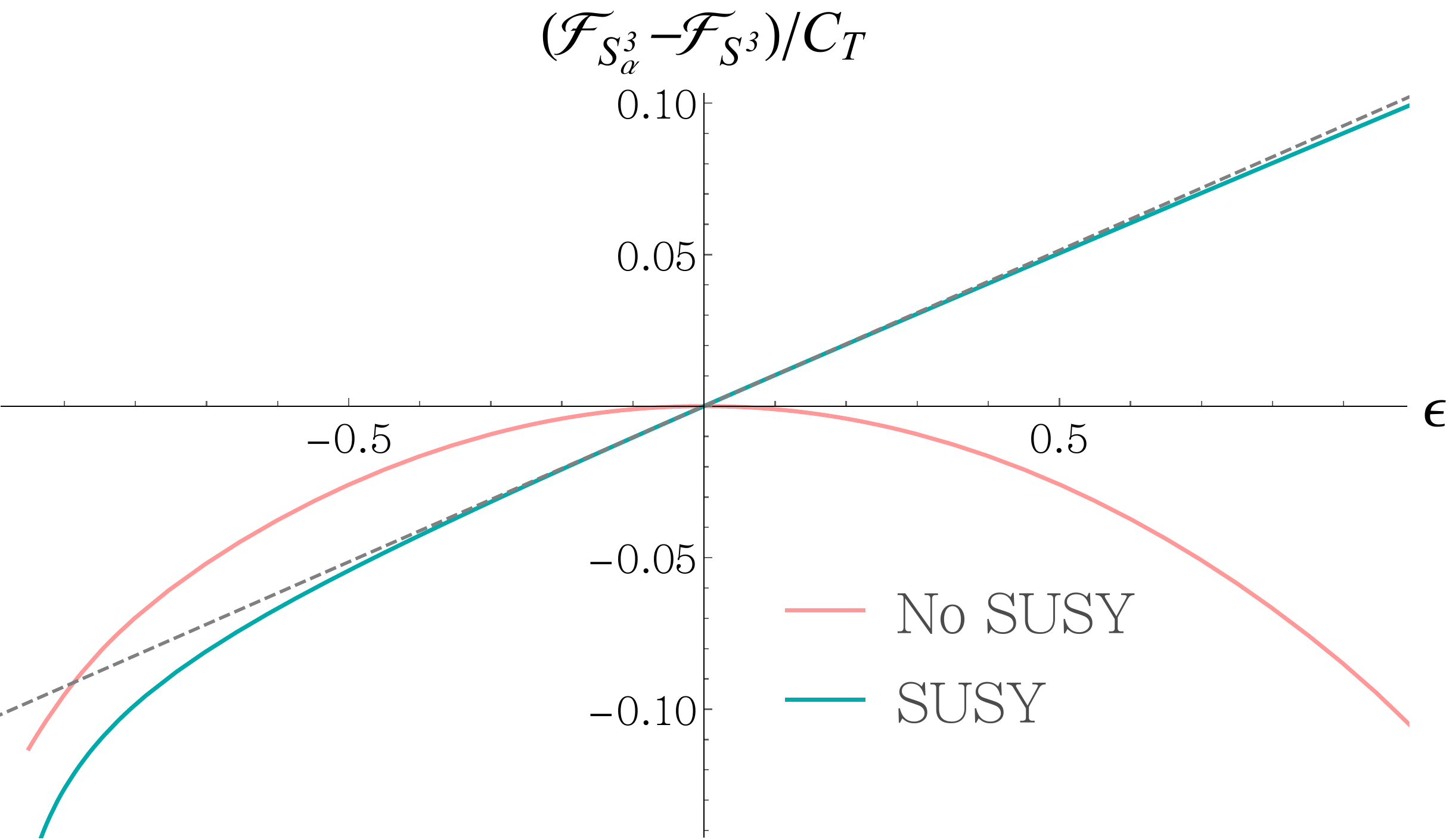}
	\caption{A comparison between the supersymmetric free energy of a free chiral multiplet (cyan), $\mathfrak{F}_{\rm chiral}(1/2,\epsilon) $, and the non-supersymmetric free energy of a theory of two conformally coupled real scalars and one Dirac fermion (pink). The dashed line depicts the linear holographic result in Equation \eqref{Fholosusy}.
	}\label{fig:F3susy}
\end{figure}

To illustrate this difference, we focus on the simple example of a theory of a single $\mathcal{N}=2$ chiral multiplet, $X$. The bosonic content of this multiplet is simply given by a complex scalar and corresponding complex fermion. The non-supersymmetric free energy $\mathcal{F}(\epsilon)$ for a \text{free} $\mathcal{N}=2$ chiral multiplet is thus easily obtained by combining our numerical results in Section \ref{ffc} and \ref{sec:3dferm}, and is plotted in Figure \ref{fig:F3susy}.

On the other hand, the supersymmetric free energy of a CFT involving only a chiral multiplet on the squashed sphere with metric \eqref{sq3} is computed by localization and found to depend only on the R-charge of the chiral multiplet, $\Delta$ --- see \cite{Imamura:2011wg} and \cite{Closset:2012ru,Nishioka:2013gza}. The result is compactly written as the following integral
\begin{equation}\label{F3dchiral}
\mathfrak{F}_{\rm chiral}(\Delta,b) = - \int_{0}^{\infty} \frac{dx}{2x} \left( \frac{\sinh(2(1-\Delta)\omega x)}{\sinh(bx)\sinh(x/b)} - \frac{2\omega(1-\Delta)}{x}\right)\;,
\end{equation}
where, following \cite{Nishioka:2013gza}, we have used a new definition of the squashing parameter given by the relations\footnote{We choose the branch with $b>0$.} 
\begin{equation}\label{epsbmap}
\epsilon = \omega^2-1\;, \qquad\qquad \omega \equiv \frac{b+b^{-1}}{2}\;.
\end{equation}
The integral in \eqref{F3dchiral} can be evaluated explicitly for the free chiral multiplet with $\Delta=1/2$. Using this value for $\Delta$, we have evaluated numerically the integral in \eqref{F3dchiral} as a function of $b$ (and thus $\epsilon$) and the results are plotted in Figure \ref{fig:F3susy}. We have used our usual normalization and the fact that for a free chiral multiplet one has  $\ctt=6$.  It is clear from this plot that the non-supersymmetric and supersymmetric free energies have quite different behavior as a function of the squashing parameter $\epsilon$.\footnote{For $\epsilon=0$, or equivalently $b=1$, the background R-symmetry gauge field has to vanish and the supersymmetric and non-supersymmetric free energies for the free chiral superfield become equal, $\mathfrak{F}_{\rm chiral}(1/2,1) = \mathcal{F}_{\rm chiral}(0)=\frac{\log{2}}{2}.$}

Another well-known CFT for which the integral in \eqref{F3dchiral} can be evaluated explicitly is the theory of a chiral multiplet with a holomorphic superpotential $W=X^3$. This is a supersymmetric relevant deformation that leads to an interacting fixed point in the IR, known as the critical Wess-Zumino CFT --- see, \eg \cite{Bobev:2015jxa}. In this interacting theory, the chiral operator is protected and has $\Delta=2/3$. We note in passing that $\ctt$ for the critical Wess-Zumino model can also be evaluated by localization (or via the conformal bootstrap) and one has $\ctt=\frac{32}{81}\left(16-9\sqrt{3}/\pi\right)\approx 4.36071$ \cite{Nishioka:2013gza,Bobev:2015jxa,Witczak-Krempa:2015jca}. We have evaluated numerically the dependence of the free energy $\mathfrak{F}$ on $\epsilon$ for this interacting theory, and we find that (using our normalization with $\ctt$) this function is almost indistinguishable (within the chosen resolution) from the supersymmetric curve for the free chiral multiplet in Figure \ref{fig:F3susy}. We emphasize that the two functions are not the same, as is clear from \eqref{F3dchiral}, they just take very similar numerical values for a large range of the parameter $\epsilon$.

Finally, let us mention that the supersymmetric squashed sphere partition function can also be computed for three-dimensional $\mathcal{N}=2$ SCFTs with a holographic dual --- see \cite{Imamura:2011wg,Martelli:2011fu}. Using the same conventions and notation as in Section \ref{holo} one finds
\begin{equation}
\mathfrak{F}_{\rm holo}(b) = \left(b+\frac{1}{b}\right)^2 \frac{\pi\ell^2}{8G}\;.
\end{equation}
This can be easily converted to a function of the squashing parameter $\epsilon$ used throughout this work
\begin{equation}\label{Fholosusy}
\mathfrak{F}_{\rm holo}(\epsilon) =  \mathfrak{F}_{\rm holo}(0)+\frac{\pi^2}{96} \ctt \epsilon\;.
\end{equation}
This simple linear function of $\epsilon$ is clearly very different from the quadratic holographic result in \eqref{sexy} derived for non-supersymmetric CFTs with a gravity dual. The difference should be attributed to the fact that in the supersymmetric setup one has to turn on a particular fixed combination of background fields in the QFT which then induce certain couplings in the bulk four-dimensional supergravity that modify the supergravity on-shell action.

\subsection{More general squashings}

We can also study a more general family of squashing deformations of the round $S^3$ metric given by the following metric
\begin{align}\label{doubly}
ds^2= \frac14 \left( \sigma^2_1 + \frac{1}{(1+\beta)}\sigma^2_2 +  \frac{1}{(1+\alpha)}\sigma^2_3 \right) \, ,
\end{align}
where the left-invariant $SU(2)$ 1-forms $\sigma_1$, $\sigma_2$ and $\sigma_3$ are given by
\begin{align}
\sigma_1=  \sin\theta \cos \tilde{\psi} d\phi - \sin\tilde{\psi} d\theta\, , \quad 
\sigma_2= \sin\theta \sin \tilde{\psi} d\phi + \cos\tilde{\psi} d\theta\, , \quad
\sigma_3= d\tilde{\psi}+\cos\theta d\phi\, .
\end{align}
Observe that \req{doubly} reduces to the metric of $S^3_{\alpha}$ in \req{sq3} when $\beta=0$. The metric in \eqref{doubly} preserves only an $SU(2)$ subgroup of the $SO(4)$ isometry group of the round $S^3$.

The eigenvalues of the scalar Laplacian for the metric in \eqref{doubly} can be computed numerically --- see \cite{PhysRevD.8.1048, Bobev:2016sap}. Therefore, the free energy for a conformally coupled free scalar field can be found numerically using a method similar to the one employed in Section \ref{ffc} for the single squashed sphere with $\beta=0$. The details of this analysis are presented in \cite{Bobev:2016sap} for a large range of the parameters $\alpha$ and $\beta$. Here we would like to point out some salient features of these results. First, following an argument completely analogous to the one in Section \ref{party}, one can show that the round sphere free energy is a local extremum of the function $\mathcal{F}(\alpha,\beta)$, \ie $\partial_{\alpha}\mathcal{F}|_{\alpha=\beta=0}=\partial_{\beta}\mathcal{F}|_{\alpha=\beta=0}=0$. The  numerical results in \cite{Bobev:2016sap} are in harmony with this general statement to a very good accuracy. In addition to that, we can extract the following leading order behavior of the free energy for small values of $\alpha$ and $\beta$. The numerical results are in a very good agreement with the following formula (we use $\ctt=3/2$ for the free scalar)
\begin{equation}\labell{Falphabeta}
\mathcal{F}_{S^{3}}(\alpha,\beta)=\mathcal{F}_{S^{3}}(0,0)-\frac{\pi^2\ctt}{96}(\alpha^2+\beta^2-\alpha\beta)+\mathcal{O}\left(\alpha^3,\beta^3\right)\,.
\end{equation}
Clearly this is very similar to the result in \eqref{f48} and it is natural to conjecture that \eqref{Falphabeta} is valid for general CFTs. This conjecture is supported also by the holographic results in \cite{Bobev:2016sap}. In particular, a two-parameter family of AdS-Taub-NUT solutions of the Einstein theory with a negative cosmological constant were numerically constructed in \cite{Bobev:2016sap}. These solutions have the metric \eqref{doubly} as an asymptotic boundary. Using holographic renormalization, one can extract the finite part of the on-shell gravitational action and, as in Section \ref{holo}, this is interpreted as the free energy of the dual CFT. These numerical holographic results  are again in good agreement with \eqref{Falphabeta}. Finally, we would like to emphasize that if \eqref{Falphabeta} holds for general CFTs, it implies that, for unitary theories, the round sphere free energy is a local maximum of $\mathcal{F}(\alpha,\beta)$ in the space of squashing deformations given by \eqref{doubly}.

\subsection{Connection to entanglement entropy}
\label{EESD}

In Section \ref{susti}, we pointed out the similarity between our results for the dependence of $\mathcal{F}_{S_{\alpha}^3}$ on the squashing parameter, and those corresponding to the universal contribution to the entanglement entropy of corner regions in three-dimensional CFTs. In both situations, $\ctt$ controls the leading (quadratic) correction to the most symmetric configuration (round sphere and absence of corner respectively), and in both, the curves corresponding to very different theories collapse to an almost-universal curve when normalized by this quantity. The connection to entanglement entropy results can, in fact, be made more precise, as we explain now.

On one hand, as shown in \cite{Casini:2011kv,Dowker:2010yj}, the universal contribution to the entanglement entropy of a region bounded by a $(d-2)$-sphere is equivalent to the free energy of the corresponding theory on $S^d$ for any odd-dimensional CFT, namely
\begin{equation}\label{chm}
s_{\rm \ssc EE}(S^{d-2})=-\mathcal{F}_{S^{d}}\, , \quad \text{for odd}\,\, d\, .
\end{equation}
On the other hand, a lot of effort has been invested recently in understanding how the entanglement entropy of planar or spherical entangling regions changes under small geometric deformations --- see \cite{Allais:2014ata,Mezei14,Rosenhaus:2014zza,Rosenhaus:2014woa,Lewkowycz:2014jia,Bueno1,Bueno2,Miao2015,Bueno4,faulkner15,Bianchi:2015liz,Bueno5,Carmi:2015dla,Dong:2016wcf,Bianchi:2016xvf,Balakrishnan:2016ttg,Fonda:2015nma} and references therein. The result of main interest for us (proven in full generality in \cite{faulkner15}) states that, given a general CFT in arbitrary (odd or even) dimensions, if we consider a deformed planar or spherical entangling surface, the leading correction to the entanglement entropy universal term  is quadratic in the deformation (schematically)
\begin{equation}\labell{s00}
s_{\rm \ssc EE}=s_{\rm \ssc EE}^{(0)}+\frac{\epsilon^2}{2}\,s_{\rm \ssc EE}^{(2)}+\mathcal{O}(\epsilon^3)\, ,
\end{equation}
and proportional to the central charge, $\ctt$, controlling the stress-tensor two-point function, \ie
\begin{equation}\labell{s20}
s_{\rm \ssc EE}^{(2)}\propto  \ctt\, ,
\end{equation}
where the proportionality constant depends on the explicit details of the geometric deformation. 

The similarity between \req{s00}, \req{s20} and \req{conj2}, \req{conj2t} is manifest, and particularly striking in the case of a deformed spherical entangling surface $S^{d-2}$ and a deformed spherical background $S^d$, in which case \req{chm} implies that the leading contributions in \req{conj2} and \req{s00} exactly coincide. Note, however, that the identity \req{chm} is in principle only valid for round spheres $S^{d-2}$ and $S^d$. The fact that the leading correction to the free energy, under geometric perturbations,  in odd-dimensional spheres is also controlled by $\ctt$, suggests that the relation \req{chm} might somehow be extended to a relation between the free energy of CFTs in deformed spheres $S^d$, and the entanglement entropy of deformed $S^{d-2}$ entangling surfaces. This looks like an intriguing possibility worth further study. Note that a $d$-dimensional sphere can be deformed in more ways than a $(d-2)$-dimensional one, which means that any putative relation generalizing \req{chm} along this lines should no longer be a bijection.

Observe also that \req{s00} and \req{s20} capture, in particular, the case of an almost-smooth corner entangling region, for which the deformation parameter is naturally identified with the corner opening angle, $\epsilon= \theta-\pi$.\footnote{Note that, in that case, the universal term is not constant, but scales logarithmically with $\ell/\delta$, where $\ell$ and $\delta$ are, respectively: some relevant macroscopic scale in the entangling surface and a short-distance cutoff.} With this identification, the analogy established in Section \ref{susti} between the corner function as a function of the opening angle and the squashed-sphere as a function of the deformation parameter becomes clear.

\subsection{Outlook}

It is clear that in this work we have only scratched the surface in understanding how the partition function of odd-dimensional CFTs placed on deformed conformally flat manifolds depends on the deformations parameters. There are some clear avenues for further work that will shed more light on this issue. 

In Sections \ref{ffc}, \ref{sec:3dferm} and \ref{freee}, we computed the squashed sphere free energy of free theories numerically. It will be desirable to find an analytic method to compute these partition functions. The zeta-function approach employed in \cite{Hartnoll:2005yc} might be a good line of attack. It will also be interesting to extend the five-dimensional results of Section \ref{freee} to free Dirac fermions. To this end one will need to find the eigenvalues of the Dirac operator on the squashed sphere in \eqref{squashedSd} generalizing the results of \cite{Gibbons198098,Dowker:1998pi}. It will of course be most welcome if we have tools to access this squashed sphere free energy for interacting non-supersymmetric CFTs. These explicit calculations will allow us to gain further insight into squashed sphere partition functions and will also serve as a test ground for the various conjectures and speculations we made in this work. In particular, it would be interesting to understand whether the speculations about a possible extension of the Hofman-Maldacena type bounds for CFTs on squashed sphere discussed in Section \ref{susti} have any merit. Perhaps a more modest goal will be to find a way to compute analytically the third order correction to the free energy in the small $\epsilon$ expansion, $\mathcal{F}'''(0)$, in \eqref{eq:Ftripplep}. The numerical results for free theories in three dimensions suggest that this coefficient is a very small number and it is desirable to understand why this is the case. Finally, it would be very interesting to make the tantalizing connections with entanglement entropy for non-spherical regions that we outlined in Section \ref{EESD} more precise, and to understand if there is a generalized version of the relation \req{chm} between partition functions and entanglement entropies in odd dimensions for deformed spheres.

\section*{Acknowledgments}  

We are grateful to Louise Anderson, Marco Baggio, Adam Bzowski, Sheer El-Showk, Fri\dh rik Gautason, Thomas Hertog, Diego Hofman, Zohar Komargodski, Edoardo Lauria, Paul McFadden, Marco Meineri, M\'ark Mezei, Rob Myers, Miguel Paulos, Balt van Rees, Carlos S. Shahbazi, Gerben Venken, and Yifan Wang for interesting discussions. The work of NB is supported in part by the starting grant BOF/STG/14/032 from KU Leuven and by an Odysseus grant G0F9516N from the FWO. The work of PB is supported by a postdoctoral fellowship from the Fund for Scientific Research - Flanders (FWO). YV is supported by the European Research Council grant no. ERC-2013-CoG 616732 HoloQosmos. All of us are also supported by the KU Leuven C1 grant ZKD1118 C16/16/005, by the Belgian Federal Science Policy Office through the Inter-University Attraction Pole P7/37, and by the COST Action MP1210 The String Theory Universe. NB would like to thank the members of the theory group at CERN for the warm hospitality in the final stages of this work.

\appendix

\section{3-point function of $T_{\mu\nu}$}
\label{App:T3pt}

The 2-point function of the energy-momentum tensor is presented in \eqref{t2p} and \eqref{Iabcddef}. Here we summarize the known expression for the 3-point function of the energy-momentum tensor of CFTs in general dimension \cite{Osborn} (see also \cite{Erdmenger:1996yc,Hofman:2008ar,Buchel:2009sk}). In Euclidean flat space the 3-point function is
\begin{equation}
\braket{T_{ab}(X)T_{cd}(Y)T_{ef}(Z)}_{\mathbb{R}^d}=\frac{\mathcal{I}_{ab,a'b'}(X-Z)\mathcal{I}_{cd,c'd'}(Y-Z)}{(X-Z)^{2d}(Y-Z)^{2d}}\, t_{a'b',c'd',ef}(W)\, ,
\end{equation}
where we have defined
\begin{equation}
\begin{split}
	\mathcal{I}_{ab,cd}(X) &\equiv\left(\delta_{ae}-\frac{2X_a X_e}{X^2}\right) \left(\delta_{bf}-\frac{2X_b X_f}{X^2}\right)\mathcal{E}_{ef, cd}\;,\\
	\mathcal{E}_{ab,cd} &\equiv\frac{1}{2}(\delta_{ac}\delta_{bd}+\delta_{ad}\delta_{bc})-\frac{1}{d}\delta_{ab}\delta_{cd}\;,\\
	W_{a} &\equiv\frac{(X-Z)_a}{(X-Z)^2}-\frac{(Y-Z)_a}{(Y-Z)^2}\, , \qquad W^2\equiv\frac{(X-Y)^2}{(X-Z)^2(Y-Z)^2}\,.
\end{split}
\end{equation}
The tensor $t_{a'b',c'd',ef}(W)$ is given by the unwieldy expression
\begin{align}\label{ititis}
&t_{ab,cd,ef}(W)\equiv\,\frac{\mathcal{A}}{(W^2)^{d/2}} \mathcal{E}_{ab,gh} \mathcal{E}_{cd,hn} \mathcal{E}_{ef,ng}\\ \notag 
&+(\mathcal{B}-2\mathcal{A}) \, \mathcal{E}_{ef,gh} \mathcal{E}_{cd,hn} \mathcal{E}_{ab,mg}\frac{W_n W_m}{(W^2)^{\frac{d+1}{2}}}\\ \notag 
&-\mathcal{B}\, \left(\mathcal{E}_{ab,gh} \mathcal{E}_{cd,hn} \mathcal{E}_{ef,mg}+(ab)\leftrightarrow (cd)\right)\frac{W_n W_m}{(W^2)^{\frac{d+1}{2}}}\\ \notag 
&+\frac{\mathcal{C}}{(W^2)^{d/2}}\, \left(\mathcal{E}_{ab,cd}\left(\frac{W_eW_f}{W^2}-\frac{1}{d}\delta_{ef} \right)+  \mathcal{E}_{cd,ef}\left(\frac{W_aW_b}{W^2}-\frac{1}{d}\delta_{ab} \right)+
 \mathcal{E}_{ef,ab}\left(\frac{W_cW_d}{W^2}-\frac{1}{d}\delta_{cd} \right)
\right)\\ \notag
&+(\mathcal{D}-4\mathcal{C})\mathcal{E}_{ab,mn} \mathcal{E}_{cd,mp} \left(\frac{W_eW_f}{W^2}-\frac{1}{d}\delta_{ef}  \right)\frac{W_n W_p}{(W^2)^{\frac{d+1}{2}}}\\ \notag 
&-(\mathcal{D}-2\mathcal{B}) \left(\mathcal{E}_{cd,mn} \mathcal{E}_{ef,mp} \left(\frac{W_aW_b}{W^2}-\frac{1}{d}\delta_{ab}  \right)+ (ab)\leftrightarrow (cd) \right)\frac{W_n W_p}{(W^2)^{\frac{d+1}{2}}}\\ \notag
&+\frac{(\mathcal{E}+4\mathcal{C}-2\mathcal{D})}{(W^2)^{d/2}} \left(\frac{W_aW_b}{W^2}-\frac{1}{d}\delta_{ab}  \right)\left(\frac{W_cW_d}{W^2}-\frac{1}{d}\delta_{cd}  \right)\left(\frac{W_eW_f}{W^2}-\frac{1}{d}\delta_{ef}  \right)\,,
\end{align}
where $(\mathcal{A},\mathcal{B},\mathcal{C},\mathcal{D},\mathcal{E})$ are constants. Only three of these constants are independent. This is made manifest by the following relations 
\begin{equation}
\begin{split}
\mathcal{D}&=\frac{(d^2-4)\mathcal{A}+(d+2)\mathcal{B}-4d \mathcal{C}}{2}\, ,\\ 
\mathcal{E}&=\frac{d(d+6)\mathcal{B}-2d(d+10)\mathcal{C}+4(d^2-4)\mathcal{A}}{4}\,.
\end{split}
\end{equation}
There is an alternative parametrization of these three independent constants which is convenient for the analysis of unitarity bounds in \cite{Hofman:2008ar}
\begin{equation}
\begin{split}
\mathcal{A}&=2\ctt d^2\frac{(d(1-d^2)+(d+1)^2t_2+(3d+1)t_4)}{(d-1)^3(d+1)^2\mathcal{S}_{d-1}^3}\, ,\\ 
\mathcal{B}&=\ctt  d\frac{(2(1+d^2(-2+d+d^2-d^3))+(3+4d+d^4)t_2+2(2+2d-d^2+d^3)t_4)}{2(d-1)^3(d+1)^2\mathcal{S}_{d-1}^3}\,,\\
\mathcal{C}&=\ctt  d \frac{(d(d-1)(d+1)(1-2d(d-1))+(d+1)(d^3-3)t_2-(4-d(d-1)(3d+2))t_4)}{2(d-1)^3(d+1)^2\mathcal{S}_{d-1}^3}\,,
\end{split}
\end{equation}
where $\mathcal{S}_{d-1}$ was defined below \eqref{t2p}. 

The important upshot from these explicit formulae is that the 3-point function of the energy-momentum tensor of CFTs in general dimension $d$ is completely determined by conformal symmetry up to three real constants --- $\ctt$, $t_2$, and $t_4$. The constant $\ctt$ is the same as the one appearing in the 2-point function of the energy-momentum tensor \eqref{t2p}. For low values of the dimension $d$, there are kinematical accidents that further reduce the number of independent constants. For example, in $d=3$ one finds that $t_2=0$, which in turn implies that \eg $\mathcal{A}$ can be written in terms of $\mathcal{B}$ and $\mathcal{C}$.

Finally let us point out that the 3-point function of the energy-momentum tensor on a conformally flat manifold can be expressed in terms of the correlator in flat space using the expression
\begin{equation}
\braket{T_{\mu\nu}(x)T_{\rho\sigma}(y)T_{\gamma\delta}(z)}_{\mathcal{M}}=\Omega^{d-2}(x)\Omega^{d-2}(y)\Omega^{d-2}(z) M_{\mu\nu}^{ab}M_{\rho\sigma}^{cd}M_{\gamma\delta}^{ef}\braket{T_{ab}(X)T_{cd}(Y)T_{ef}(Z)}_{\mathbb{R}^d}\,,
\end{equation}
where $\Omega(x)$ and $M_{\mu\nu}^{ab}$ are defined in \eqref{conj121} and \eqref{tens}.

\section{AdS-Taub-NUT solutions in Gauss-Bonnet gravity}
\label{TNn}

An important part in our analysis is played by the so-called Euclidean AdS-Taub-NUT solution of four-dimensional general relativity with a negative cosmological and its generalizations to higher-dimensions --- see \cite{Hawking:1998ct,Emparan:1999pm} and references therein. The metric on this space can be thought of as an $S^1$ fibration over a $(d-1)$-dimensional base space, $\mathcal{B}$, equipped with an Einstein-K\"ahler metric $g_{\mathcal{B}}$. The $(d+1)$-dimensional metric can be written in general as
\begin{equation}\label{TN1}
ds^2 = F(r)\left(d\tau + n A\right)^2+\frac{dr^2}{F(r)}+(r^2-n^2)ds_{\mathcal{B}}^2\;,
\end{equation}
where $n$ is the so-called NUT charge, $\tau$ is the coordinate of the fibre $S^1$ and $A$ is related to the K\"ahler form $J$ on $\mathcal{B}$ through $J=dA$. When the dimension of the set of fixed points of the $S^1$ isometry is less than $(d-1)$, the solution describes a NUT, while when it is equal to $(d-1)$ it describes a Bolt.\footnote{This terminology was introduced in \cite{Gibbons:1979xm}.} The Bolt solutions are not smoothly connected to the $AdS_{d+1}$ metric and will not be considered in this work. The explicit form of the function $F(r)$ depends on the dimension $d$ and on the particular gravitational theory of interest.

Here we focus on metrics for which the base space is the complex projective plane, \ie $\mathcal{B}=\mathbb{CP}^{\frac{d-1}{2}}$. We can write the metric on that space in the following recursive form (we use the notation $k\equiv \frac{d-1}{2}$)
\begin{equation}\label{TNcp}
ds^2_{\mathbb{CP}^{k}} = (2k+2)\left\{d\xi^2_k+\sin^2\xi_k \cos^2\xi_k \left(d\psi_k+\frac{1}{2k}A_{\mathbb{CP}^{k-1}}\right)^2+\frac{1}{2k}\sin^2\xi_k ds^2_{\mathbb{CP}^{k-1}} \right\}\;,
\end{equation}
where $A_{\mathbb{CP}^{k-1}}$ is related to the K\"ahler form of $\mathbb{CP}^{k-1}$ as $J_{\mathbb{CP}^{k-1}}=dA_{\mathbb{CP}^{k-1}}$ and $\xi_k$ and $\psi_k$ are the new coordinates of $\mathbb{CP}^{k}$ with respect to $\mathbb{CP}^{k-1}$. The normalization of \req{TNcp} is chosen such that the Ricci tensor obeys $R_{ab}=g_{ab}$. The 1-form  $A_{\mathbb{CP}^{k}}$ is given by
\begin{equation}\label{acp}
A_{\mathbb{CP}^{k}} = (2k+2)\sin^2\xi_k\left(d\psi_k+\frac{1}{2k}A_{\mathbb{CP}^{k-1}}\right)\;.
\end{equation}
It is clear from these formulae that it is possible to reconstruct the K\"ahler form and the metric of $\mathbb{CP}^{k}$ from the analogous information for $\mathbb{CP}^{k-1}$ in a recursive manner starting from the $k=1$ case corresponding to the usual round metric on $S^2$
\begin{equation}\label{cp11}
\begin{split}
ds^2_{\mathbb{CP}^{1}} &= 4\left\{d\xi^2_1+\sin^2\xi_1 \cos^2\xi_1 d\psi_1^2\right\}\,,\\ 
A_{\mathbb{CP}^{1}} &= 4\sin^2\xi_1 d\psi_1\,.
\end{split}
\end{equation}
While NUT and Bolt solutions in various dimensions have been constructed in the literature for many classes of theories, our interest here is in even-dimensional Gauss-Bonnet gravities, whose Euclidean action is given by 
\begin{equation}\label{gbd}
I_{\rm GB}= -\frac{1}{16\pi G}\int_{\mathcal{M}} d^{d+1}x\, \sqrt{g} \left[\frac{d(d-1)}{\ell^2}+R+\frac{\ell^2 \lambda_{\rm GB}\,\mathcal{X}_4}{(d-2)(d-3)} \right]\, ,
\end{equation}
where $G$ is the Newton constant, $\ell$ sets the scale of the cosmological constant, $\lambda_{\rm GB}$ is the Gauss-Bonnet coupling and $\mathcal{X}_4$ is the Euler density defined below \eqref{gb}. The not very illuminating form of $F(r)$ for the NUT solutions with a $\mathbb{CP}^{\frac{d-1}{2}}$ base space for general even-dimensional Gauss-Bonnet gravity can be found in \cite{Dehghani:2005zm} (see also references therein). For our purposes, it suffices to point out that the asymptotic metric of such solutions can be recast in the form
\begin{equation}\label{TNbd}
ds^2 =\frac{\ell^2 }{r^2} dr^2+r^2 ds^2_{\ssc \rm bdy.}\, ,
\end{equation}
where the boundary metric reads
\begin{equation}\label{TNnn}
ds^2_{\ssc \rm bdy.}=\frac{1}{(d+1)} ds^2_{\mathbb{CP}^{k}}+\left(\frac{n^2 (d+1)f_{\infty}}{\ell^2}\right) \left(d\psi + \frac{A_{\mathbb{CP}^{k}}}{(d+1)}\right)^2\, \, ,
\end{equation}
and where we made the coordinate redefinitions $\tau \rightarrow \psi \cdot n (d+1)$ and $r\rightarrow r/\sqrt{d+1}$ and defined $f_{\infty}\equiv 2/(1+\sqrt{1-4\lambda_{\rm GB}})$.

When
\begin{equation}\label{nl}
\frac{n^2}{\ell^2}=\frac{1}{(d+1)f_{\infty}}\, ,
\end{equation}
the boundary metric becomes the one corresponding to a usual round sphere $S^d$ with $R_{ij}=(d-1)g_{ij}$, and \req{TNbd} becomes the metric of pure AdS$_{d+1}$. For other values of the NUT charge, \req{TNnn} is precisely the metric of a squashed sphere $S^d_{\alpha}$. Comparing \req{TNbd} and \req{TNnn} with \req{squashedSd} one finds the following relation between $n$ and the squashing parameter $\alpha$
 \begin{equation}\label{nalpharel}
  \frac{n^2}{\ell^2}=\frac{1}{(d+1)f_{\infty}(1+\alpha)}.
 \end{equation}
 %
 
 \subsection{Holographic free energy}

To make the variational problem associated with the functional \req{gbd} well-defined one has to add the following boundary term \cite{Gibbons:1976ue,Myers:1987yn,Teitelboim:1987zz}
\begin{equation}\label{gbdbd}
I_{\rm GB,\, bdy.}= -\frac{1}{8\pi G}\int_{\mathcal{\partial M}}d^{d}x\, \sqrt{h} \left[K +\frac{2\ell^2 \lambda_{\rm GB}}{(d-2)(d-3)} \left[\mathcal{J}-2 \mathcal{G}_{ij}K^{ij} \right]\right]\, ,
\end{equation}
where $h$ is the determinant of the induced metric $h_{ij}$ on the boundary,\footnote{This is defined as $h_{ij}=t^i_{\mu}t^j_{\nu}g^{\mu\nu}$, where $t^i_{\mu}=\partial y^i/\partial x^{\mu}$ and the $y^i$ are coordinates on $\partial {\mathcal{M}}$. Similarly, $K_{ij}=t^i_{\mu}t^j_{\nu}K^{\mu\nu}$, where $K_{\mu\nu}=\nabla_{\mu}n_{\nu}$, with $n_{\nu}$ the unit vector orthogonal to the boundary.} $K_{ij}$ is the second fundamental form associated to the vector normal to $\partial M$, $K$ its trace, $\mathcal{G}_{ij}$ the Einstein tensor associated to $h_{ij}$ and $\mathcal{J}$ the trace of the boundary tensor
\begin{equation}\label{J}
\mathcal{J}_{ij}=\frac{1}{3}\left(2K K_{ik}K^{k}_{j}+K_{kl}K^{kl}K_{ij}-2K_{ik}K^{kl}K_{lj}-K^2 K_{ij} \right)\, .
\end{equation}
For holographic applications, one would like to have solutions to this Gauss-Bonnet gravity theory with a finite on-shell action. To achieve this, we can rely on the standard techniques of holographic renormalization \cite{Skenderis:2002wp} and supplement the original action \req{gbd} with the following counterterms \cite{Emparan:1999pm,Mann:1999pc,Balasubramanian:1999re,Brihaye:2008xu}
\begin{align}\label{cttt}
 I_{\rm GB,\, ct.}&= -\frac{1}{8\pi G}\int_{\mathcal{\partial M}}d^{d}x\, \sqrt{h} \left\{-\frac{(d-1)(f_{\infty}+2)}{3\ell f_{\infty}^{1/2}}-\frac{\ell (3f_{\infty}-2)\Theta[d-3]}{2f_{\infty}^{3/2}(d-2)}\mathcal{R} \right. \\ \notag   &\left. -\frac{\ell^3\Theta[d-5]}{2 f_{\infty}^{5/2}(d-2)^2(d-4)}\left[(2-f_{\infty})\left(\mathcal{R}_{ij}\mathcal{R}^{ij}-\frac{d}{4(d-1)}\mathcal{R}^2 \right)-\frac{(d-2)(1-f_{\infty})}{(d-3)} \mathcal{X}_4^{(h)}\right]+\dots\, 
 \right\}
\end{align}
where we have included all the relevant terms for $d<7$ theories and where $\Theta[x]=1$ if $x\geq 0$ and zero otherwise. The terms indicated by dots in \eqref{cttt} are proportional to $\Theta[d-m]$ with $m\geq  7$ and thus vanish for $d\leq6$, which is of relevance for our work. Note that $\mathcal{R}_{ij}$ and $\mathcal{R}$ are the Ricci tensor and scalar associated to $h_{ij}$ and $
\mathcal{X}_4^{(h)}=\mathcal{R}-4\mathcal{R}_{ij}\mathcal{R}^{ij}+\mathcal{R}_{ijkl}\mathcal{R}^{ijkl}\, ,
$
is the Gauss-Bonnet density associated to this metric. The final result for the regularized action takes the form
\begin{equation}\labell{osa}
 I_{\rm GB,\, finite}=I_{\rm GB}+I_{\rm GB,\, bdy.}+ I_{\rm GB,\, ct.}\, .
\end{equation}
%

\subsubsection*{Four-dimensional Einstein gravity}
Let us consider the case of four-dimensional Einstein gravity. Since the Gauss-Bonnet term is topological in four-dimensions we can set $f_{\infty}\rightarrow 1$ ($\lambda_{\rm GB}\rightarrow 0$)  and $d=3$ in \req{gbd}, \req{gbdbd} and \req{cttt} respectively. The AdS-Taub-NUT solution of this theory was constructed in \cite{Chamblin:1998pz,Hawking:1998ct} and the corresponding $F(r)$ reads 
\begin{equation}\label{TN1n}
F(r)=\frac{(r^2+n^2)-2mr+\ell^{-2}(r^4-6n^2r^2-3n^4)}{r^2-n^2}\, ,
\end{equation}
where $m=n-4n^3/\ell^2$ is the mass of the NUT. The free energy of this solution can be obtained using \req{osa}, and one finds \cite{Emparan:1999pm}
\begin{equation}\label{tnf}
\mathcal{F}_{\textrm{E},\,S^3_{\alpha}}=\frac{\pi \ell^2}{2G}\frac{(1+2\alpha)}{(1+\alpha)^2}\, .
\end{equation}
Here we have used the relation \eqref{nalpharel} to express the NUT charge $n$ in terms of the squashing parameter $\alpha$.
 
\subsubsection*{Six-dimensional Gauss-Bonnet gravity} 
The six-dimensional theory is more involved, since the Gauss-Bonnet term plays a non-trivial role. The AdS-Taub-NUT solution with base space $\mathbb{CP}^2$ is given by \req{TN1}, with \cite{Dehghani:2005zm}
\begin{equation}\label{TN1ngb}
F(r)=\frac{(r^2-n^2)^2}{2\ell^2\lambda_{\rm GB}(r^2+n^2)}\left[1+\frac{2\ell^2\lambda_{\rm GB}}{3(r^2-n^2)}-\sqrt{1+2\lambda_{\rm GB}\ell^2L(r)} \right]\, ,
\end{equation}
and
\begin{equation}\label{TN1ngb}
L(r)=\frac{\ell^2(12 n^2-2\ell^2\lambda_{\rm GB})(r^4+6r^2n^2+n^4)+(9 m r \ell^2 -18(r^6-5 n^2 r^4 +15 n^4 r^2+5 n^6)) (r^2+n^2)}{9\ell^2(r^2-n^2)^4}  \, .
\end{equation}
Using \req{osa}, we obtain the following result for the free energy of this solution
\begin{equation}\label{tnfgb22}
\mathcal{F}_{\textrm{GB},\, S^{5}_{\alpha}}= -\frac{  \pi ^2 \ell^4\left[6\alpha ^2 (f_{\infty}-1)+(3\alpha+1)(3
	f_{\infty}-4)\right]}{3G (\alpha+1)^3 f_{\infty}^3 }\;.
\end{equation}
As already stressed in the main text, this free energy was originally computed in \cite{Clarkson:2002uj} in the particular case of Einstein gravity ($\lambda_{\rm GB}=0$), and in \cite{KhodamMohammadi:2008fh} for the full Gauss-Bonnet theory. Both calculations disagree from ours by an overall factor of $9/8$ in the free energy. The fact that our result \req{tnfgb22} satisfies the general relation \req{f485}, which was derived in field theory for general CFTs, gives us confidence that the expression in \req{tnfgb22} is correct.

\section{Conformal Laplace operator on squashed spheres}
\label{ssgd}

In this appendix, we construct the eigenvalues of the conformal Laplacian on a general-dimension squashed sphere with the metric \req{squashedSd}.
Given an arbitrary Riemannian manifold, the conformal Laplace operator is defined as
\begin{equation}
\mathcal{D} = -\nabla^2 +\frac{d-2}{4(d-1)} R\;,
\end{equation}
where $R$ and $\nabla^2$ are the Ricci scalar and the usual Laplace operator respectively.

The eigenvalues of $\mathcal{D}$ and their corresponding degeneracies on the round unit-radius sphere $S^d$ are respectively given by (see for example \cite{Klebanov:2011gs})
\begin{align}
\lambda_{d,\ell}=\ell(\ell+d-1)+\frac{d(d-2)}{4}\, , \qquad m_{d,\ell}=(2\ell+d-1)\frac{(\ell+d-2)!}{(d-1)!\ell!}\, ,\label{eigenround}
\end{align}
where $\ell \geq 0$.
Now, the Laplacian on the squashed-sphere $S_{\alpha}^d$ can be written in terms of the round-sphere Laplacian as
\begin{align}
\nabla^2_{\alpha} = \nabla^2 -\alpha \partial_{\psi}^2\, .\label{Lapsq}
\end{align}
Using the expression for the Ricci scalar associated to \req{squashedSd},
\begin{align}
R_{\alpha} =\frac{(d-1)(d+\alpha(d+1))}{1+\alpha} \, ,\label{Riccisq}
\end{align}
it is possible to show (see below) that the eigenvalues of $\mathcal{D}$ on $S^d_{\alpha}$ read
\begin{align}
\lambda_{d,\ell,q}=\ell(\ell+d-1)+\alpha(\ell-2q)^2+\frac{(d-2)(d+(d+1)\alpha)}{4(1+\alpha)}\, ,\label{eigensq}
\end{align}
where $\ell\geq0$ and $0\leq q\leq \ell$. Their degeneracy is in turn
\begin{align}
m_{d,\ell,q}=\frac{\left(\ell+\frac{d-1}{2}\right)\prod\limits_{j=1}^{\frac{d-3}{2}}(q+j)(\ell-q+j)}{\left(\frac{d-3}{2}\right)!\left(\frac{d-1}{2}\right)!}\, .\label{degesq}
\end{align}
As a non-trivial consistency check of this formula, it is easy to show that
\begin{equation}
\sum_{q=0}^{\ell} m_{d,\ell,q} = m_{d,\ell}\,,
\end{equation}
as expected.
Similarly, one also finds that $\lambda_{d,\ell,q}=\lambda_{d,\ell}$ for $\alpha=0$.

The expressions in \req{eigensq} and \req{degesq} can be deduced by using the fact that the squashed sphere we are considering preserves the isometries of $\mathbb{CP}^{\frac{d-1}{2}}$, which in turn is a coset space
\begin{equation}
\mathbb{CP}^{\frac{d-1}{2}} \cong \dfrac{SU(\frac{d+1}{2})}{SU(\frac{d-1}{2})\times U(1)}\;.
\end{equation}
On the round sphere, the eigenvalue of the Laplacian $\lambda_{d,\ell}$ in \eqref{eigenround} is in the representation of $SO(d+1)$ with Dynkin label $(\ell,0,\ldots,0)$. The squashed sphere we are considering preserves only an $SU(\frac{d+1}{2})\times U(1)$ subgroup of $SO(d+1)$. This in turn implies that we should decompose the $(\ell,0,\ldots,0)$ representation of $SO(d+1)$ in irreducible representations of $SU(\frac{d+1}{2})\times U(1)$. Performing this decomposition in detail one finds the results in \eqref{eigensq} and \eqref{degesq}.

To illustrate this procedure, let us work out explicitly some examples in five dimensions, \ie branching of representations of $SO(6)$ under $SU(3)$. For the lowest dimensional representations of $SO(6)$ one finds (see for example \cite{Slansky:1981yr})
\begin{eqnarray}\label{SU3branching}
\mathbf{1} &=& (0,0,0) \to (0,0) = \mathbf{1}\;, \notag\\
\mathbf{6} &=& (1,0,0) \to (1,0)\oplus (0,1) = \mathbf{3}\oplus \mathbf{\bar{3}}\;, \notag\\
\mathbf{20} &=& (2,0,0) \to (2,0)\oplus(1,1)\oplus (0,2) = \mathbf{6}\oplus\mathbf{8}\oplus \mathbf{\bar{6}}\;, \\
\mathbf{50} &=& (3,0,0) \to (3,0)\oplus(2,1)\oplus (1,2)\oplus (0,3) = \mathbf{10}\oplus\mathbf{15}\oplus\mathbf{\overline{15}}\oplus \mathbf{\overline{10}}\;, \notag\\
\mathbf{105} &=& (4,0,0) \to (4,0)\oplus(3,1)\oplus(2,2)\oplus(1,3)\oplus (0,4) = \mathbf{15'}\oplus\mathbf{24}\oplus\mathbf{27}\oplus \mathbf{\overline{24}}\oplus \mathbf{15'}\;, \notag\\
\mathbf{196} &=& (5,0,0) \to (5,0)\oplus(4,1)\oplus(3,2)\oplus(2,3)\oplus(1,4)\oplus (0,5) = \mathbf{21}\oplus\mathbf{35}\oplus \mathbf{42}\oplus \mathbf{\overline{42}}\oplus \mathbf{\overline{35}}\oplus \mathbf{\overline{21}}\;.\notag
\end{eqnarray}
The numbers on the left hand side in \eqref{SU3branching} are the degeneracies on the round five-sphere \eqref{eigenround}
\begin{equation}
m_{5,0} = 1\;, ~~~ m_{5,1} = 6\;, ~~~ m_{5,2} = 20\;, ~~~ m_{5,3} = 50\;, ~~~ m_{5,4} = 105\;, ~~~ m_{5,5} = 196\;.
\end{equation}
The numbers on the right hand side in \eqref{SU3branching} are the degeneracies on the squashed five-sphere as written in \eqref{degesq}
\begin{equation}
\begin{split}
m_{5,0,0} &= 1\;, ~~~ m_{5,1,1} = 3\;, ~~~ m_{5,1,0} = 3\;, ~~~ m_{5,2,2} = 6\;, ~~~ m_{5,2,1} = 8\;, ~~~ m_{5,2,0} = 6\;,\\
m_{5,3,3} &= 10\;, ~~~ m_{5,3,2} = 15\;, ~~~ m_{5,3,1} = 15\;, ~~~ m_{5,3,0} = 10\;, \\
m_{5,4,4} &= 15\;, ~~~ m_{5,4,3} = 24\;, ~~~ m_{5,4,2} = 27\;, ~~~ m_{5,4,1} = 24\;,~~~ m_{5,4,0} = 15\;, \\
m_{5,5,5} &= 21\;, ~~~ m_{5,5,4} = 35\;, ~~~ m_{5,5,3} = 42\;, ~~~ m_{5,5,2} = 42\;,~~~ m_{5,5,1} = 35\;, ~~~ m_{5,5,0} = 21\;.
\end{split}
\end{equation}
The general rule for decomposing the $(\ell,0,\ldots,0)$ representation of $SO(d+1)$ is the following. First, one has to remove one of the Dynkin labels, since the rank of $SU(\frac{d+1}{2})$ is one lower than the rank of $SO(d+1)$. Then one writes all possible decompositions of the non-negative integer $\ell$ into a pair of non-negative integers $p_1$ and $p_2$ such that $\ell=p_1+p_2$. Finally, one writes all possible Dynkin labels with $p_1$ as the first entry and $p_2$ as the last with zeroes in between. One can then identify the integer $q$ in equations \eqref{eigensq} and \eqref{degesq} with $p_1$.

\section{Numerical regularization}
\label{ff}

In this appendix, we give more details on the numerical techniques utilized in the free field theory calculations throughout the paper. As already mentioned, our regularization scheme is based on the same technique as the one used in \cite{Anninos:2012ft} (see also \cite{Bobev:2016sap}). 

The free energy of the free scalar and fermion in arbitrary dimensions can be written as 
\begin{align}\label{eqn:Fdetdef}
 \mathcal{F} = (-1)^f\dfrac{1}{2^{f-1}}\log \det \left[\mathfrak{D}/\Lambda^f \right]\, ,
\end{align}
where $\mathfrak{D}$ is the conformal Laplace operator or the Dirac operator respectively, $f=1$ for fermions, $f=2$ for scalar fields, and $\Lambda$ is an energy cutoff. The free energy will, in general, be a UV divergent quantity which needs to be regularized. To this end, we use a heat-kernel type regulator \cite{Anninos:2012ft, Vassilevich:2003xt} 
\begin{align}
	\log \det\left[\mathfrak{D}/\Lambda^f\right] = \sum_i \int_{\Lambda^{-2}}^{\infty}\frac{dt}{t} e^{-t\lambda_i^{3-f}} \, ,\label{eqn:regulator}
\end{align}
where we have denoted the eigenvalues of $\mathfrak{D}$ by $\lambda_i$. This expression yields the determinant for modes whose energies are less than a ``soft'' cutoff $ \Lambda$, while cutting off the sum exponentially above this value. In particular, for $\lambda  \ll \Lambda^f$, one finds
\begin{align}
	- \int_{\Lambda^{-2}}^{\infty}\frac{dt}{t} e^{-t\lambda_i^{3-f}} =\log(\lambda_i^{3-f}/ \Lambda^{2})+\mathcal{O}(\lambda_i^{3-f}/ \Lambda^{2}) \ ,\label{eqn:upperincompleteGamma}
\end{align}
while for $\lambda  \gg  \Lambda^f$,
\begin{align}
	- \int_{\Lambda^{-2}}^{\infty}\frac{dt}{t} e^{-t\lambda_i^{3-f}} =-e^{-\lambda_i^{3-f}/ \Lambda^{2}}\left(\frac{\Lambda^{2}}{\lambda_i^{3-f} }+ \mathcal{O}\left(\frac{\Lambda^{4}}{\lambda_i^{6-2f}}\right)\right) \ .
\end{align}
The integral can now be split into two pieces, one with low energy modes (IR) and another with high energy modes (UV)
\begin{align}\label{eqn:detIRplusUV}
	\log \det\left[\mathfrak{D}/\Lambda^f \right] =  \textrm{det}_{UV} + \textrm{det}_{IR} \, ,
\end{align}
where 
\begin{equation}
\begin{split}
	\textrm{det}_{UV} &\equiv  \sum_{i}\int_{\Lambda^{-2}}^{\delta} \frac{dt}{t}  m_i e^{-t \lambda_{i}^{3-f}} \ , \\
	\textrm{det}_{IR} &\equiv \sum_{i} \int_{\delta}^{\infty} \frac{dt}{t} m_i e^{-t \lambda_{i}^{3-f}} = \sum_{i}m_i \Gamma(0,\lambda_i^{3-f} \delta)\ . \label{eqn:detUV}
\end{split}
\end{equation}
Here, $\Gamma(a,z)$ is the incomplete Euler Gamma function, $\delta$ is an arbitrary positive real number that we can change to get a better convergence, and $m_i$ is the multiplicity of the eigenvalue $\lambda_i$. The sum over the IR modes converges for large $\lambda_i$, and can therefore be done numerically if the maximum number of eigenvalues is chosen correctly.

The divergences are all contained in det$_{UV}$. These have to be controlled and subtracted. For the case of the free scalar field in three dimensions, one can apply the Euler-MacLaurin formula to the infinite sums to obtain these analytically \cite{Anninos:2012ft}. In more general cases, the Euler-MacLaurin formula does not give an estimate of the infinite sum. However, for a given value of $t$, this sum can be obtained numerically, as it converges fast enough if we take into account enough eigenvalues. The maximum number of eigenvalues that we will sum over will be called $n_{\rm max}$.  
The next step in the algorithm consists in evaluating this converging sum for different values of $t$, from a starting value $t_{\rm init}$, to the final value $\delta$, with step size $\Delta t$. With this sum evaluated for a large range of $t$'s, it is possible to do a numerical fit to the ``data points''. This gives the dependence of the sum on the energy cutoff. 

In order to know which function we need to use for the fit, it is necessary to know the behavior of the integrand as a function of $t$. Since $t \sim \Lambda^{-2}$, we know that it should have the same behavior as the on-shell action in terms of the cutoff powers --- see \eg \req{eqn:divergences}. Therefore, we perform a fit with the function
\begin{align}\label{fittdef}
	\textrm{fit}(t)= \frac{A_1}{t^{(d+1)/2}}+\frac{A_2}{t^{d/2}} + \frac{A_3}{t^{(d-1)/2}}+ \cdots  + \frac{A_{d+1}}{t^{1/2}} \, .
\end{align}
We expect all the $A_i$ with $i$ odd to vanish --- \ie to be very small numbers in our numerical calculations, so that the leading term of the fit contains the volume divergence, the subleading one the first curvature divergence etc. In each case, the coefficients that should vanish are used as a precision check of our numerical work. If these are really small, the fit can be trusted. To give a flavor of our results for the coefficients in \eqref{fittdef} in Table \ref{tbl:coeff}, we present explicitly the values for $A_i$ for different values of $\alpha$ in $d=3$ for the free fermion field. Notice that for the free fermion the eigenvalues in \eqref{fermioneigen} are labelled by the integers, and thus the sum over $i$ splits into 2 sums, one over $q$ and another over $n$ in \eqref{fermioneigen}. The multiplicity in this case is equal to $n$. To obtain the values in Table \ref{tbl:coeff}, we take for det$_{UV}$, $n_{\rm max}=3101$, and we start our fit from $t_{\rm init} = 10^{-6}$ with step size $\Delta t= 10^{-6}$. For the parameter $\delta$  we choose $\delta = 10^{-3}$. 

After the fit function in \eqref{fittdef} is obtained in this way, we have good control over the divergences in ${\rm det}_{UV}$ which we then subtract and, after that, evaluate the integral in \eqref{eqn:detUV}. The result is then used in \eqref{eqn:detIRplusUV} and \eqref{eqn:Fdetdef} to obtain the finite regularized value of the free energy.
\begin{table*}
  \centering
  \begin{tabular}{r||c|c|c|c} 
   $\alpha$  &  $A_1$ & $A_2$  & $A_3$ & $A_4$   
   \\
    \hline \hline 
{$-0.8$} & {$-1.57028\cdot 10^{-7}$}&  {$0.24790$}&  {$-0.067730$} &  {$7.34150$} \\ \hline
$-0.50 $&  {$-1.02049\cdot10^{-12}$}&  {$0.15670$}&  {$-4.414079\cdot 10^{-7}$} &  {$ -0.208839$} \\ \hline
$-0.21 $&  {$-5.97831 \cdot 10^{-17}$}&  {$0.12464$}&  {$-5.502979\cdot 10^{-11}$} &  {$ -0.227184$}\\ \hline
$0.0 $&  {$-3.96029 \cdot 10^{-18}$}&  {$0.11078$}&  {$-1.9311 \cdot 10^{-12} $}&  {$ -0.221557$}\\ \hline
$0.421 $&  {$-3.84916 \cdot 10^{-18}$}&  {$0.09293$}&  {$4.003657 \cdot 10^{-12}$} &  {$ -0.204216$} \\ \hline
$0.5 $&  {$-4.00575 \cdot 10^{-18}$}&  {$0.09045$}&  {$-2.37087 \cdot 10^{-12}$} &  {$ -0.201000$}\\ \hline
$7.982 $&  {$1.16793 \cdot 10^{-16}$}&  {$0.03696$}&  {$1.38332 \cdot 10^{-10}$} &  {$ -0.095824$}  \end{tabular}
  \caption{The coefficients of the fit function in \eqref{fittdef} for different values of $\alpha$. Notice that for values of $\alpha$ close to $-1$ the coefficient $A_2$ differs more significantly from the expected zero value and thus our numerical results are less accurate.  }
\label{tbl:coeff}   
\end{table*}    
%

\renewcommand{\leftmark}{\MakeUppercase{Bibliography}}
\phantomsection
\bibliographystyle{JHEP}
\bibliography{quartic-refs}

\providecommand{\href}[2]{#2}\begingroup\raggedright\begin{thebibliography}{10}

\bibitem{Pestun:2016zxk}
V.~Pestun et~al., \emph{{Localization techniques in quantum field theories}},
  \href{http://arxiv.org/abs/1608.02952}{{\tt 1608.02952}}.

\bibitem{Closset:2012vg}
C.~Closset, T.~T. Dumitrescu, G.~Festuccia, Z.~Komargodski and N.~Seiberg,
  \emph{{Contact Terms, Unitarity, and F-Maximization in Three-Dimensional
  Superconformal Theories}},
  \href{http://dx.doi.org/10.1007/JHEP10(2012)053}{\emph{JHEP} {\bf 10} (2012)
  053}, [\href{http://arxiv.org/abs/1205.4142}{{\tt 1205.4142}}].

\bibitem{Closset:2012vp}
C.~Closset, T.~T. Dumitrescu, G.~Festuccia, Z.~Komargodski and N.~Seiberg,
  \emph{{Comments on Chern-Simons Contact Terms in Three Dimensions}},
  \href{http://dx.doi.org/10.1007/JHEP09(2012)091}{\emph{JHEP} {\bf 09} (2012)
  091}, [\href{http://arxiv.org/abs/1206.5218}{{\tt 1206.5218}}].

\bibitem{Osborn}
H.~Osborn and A.~C. Petkou, \emph{{Implications of conformal invariance in
  field theories for general dimensions}},
  \href{http://dx.doi.org/10.1006/aphy.1994.1045}{\emph{Annals Phys.} {\bf 231}
  (1994) 311--362}, [\href{http://arxiv.org/abs/hep-th/9307010}{{\tt
  hep-th/9307010}}].

\bibitem{Bobev:2015jxa}
N.~Bobev, S.~El-Showk, D.~Mazac and M.~F. Paulos, \emph{{Bootstrapping SCFTs
  with Four Supercharges}},
  \href{http://dx.doi.org/10.1007/JHEP08(2015)142}{\emph{JHEP} {\bf 08} (2015)
  142}, [\href{http://arxiv.org/abs/1503.02081}{{\tt 1503.02081}}].

\bibitem{Buchel:2009sk}
A.~Buchel, J.~Escobedo, R.~C. Myers, M.~F. Paulos, A.~Sinha and M.~Smolkin,
  \emph{{Holographic GB gravity in arbitrary dimensions}},
  \href{http://dx.doi.org/10.1007/JHEP03(2010)111}{\emph{JHEP} {\bf 03} (2010)
  111}, [\href{http://arxiv.org/abs/0911.4257}{{\tt 0911.4257}}].

\bibitem{Klebanov:2011gs}
I.~R. Klebanov, S.~S. Pufu and B.~R. Safdi, \emph{{F-Theorem without
  Supersymmetry}}, \href{http://dx.doi.org/10.1007/JHEP10(2011)038}{\emph{JHEP}
  {\bf 10} (2011) 038}, [\href{http://arxiv.org/abs/1105.4598}{{\tt
  1105.4598}}].

\bibitem{Casini:2012ei}
H.~Casini and M.~Huerta, \emph{{On the RG running of the entanglement entropy
  of a circle}},
  \href{http://dx.doi.org/10.1103/PhysRevD.85.125016}{\emph{Phys. Rev.} {\bf
  D85} (2012) 125016}, [\href{http://arxiv.org/abs/1202.5650}{{\tt
  1202.5650}}].

\bibitem{Pufu:2016zxm}
S.~S. Pufu, \emph{{The F-Theorem and F-Maximization}},  2016.
\newblock \href{http://arxiv.org/abs/1608.02960}{{\tt 1608.02960}}.

\bibitem{Anninos:2012ft}
D.~Anninos, F.~Denef and D.~Harlow, \emph{{Wave function of Vasiliev's
  universe: A few slices thereof}},
  \href{http://dx.doi.org/10.1103/PhysRevD.88.084049}{\emph{Phys. Rev.} {\bf
  D88} (2013) 084049}, [\href{http://arxiv.org/abs/1207.5517}{{\tt
  1207.5517}}].

\bibitem{Anninos:2013rza}
D.~Anninos, F.~Denef, G.~Konstantinidis and E.~Shaghoulian, \emph{{Higher Spin
  de Sitter Holography from Functional Determinants}},
  \href{http://dx.doi.org/10.1007/JHEP02(2014)007}{\emph{JHEP} {\bf 02} (2014)
  007}, [\href{http://arxiv.org/abs/1305.6321}{{\tt 1305.6321}}].

\bibitem{Bobev:2016sap}
N.~Bobev, T.~Hertog and Y.~Vreys, \emph{{The NUTs and Bolts of Squashed
  Holography}}, \href{http://dx.doi.org/10.1007/JHEP11(2016)140}{\emph{JHEP}
  {\bf 11} (2016) 140}, [\href{http://arxiv.org/abs/1610.01497}{{\tt
  1610.01497}}].

\bibitem{Hawking:1998ct}
S.~W. Hawking, C.~J. Hunter and D.~N. Page, \emph{{Nut charge, anti-de Sitter
  space and entropy}},
  \href{http://dx.doi.org/10.1103/PhysRevD.59.044033}{\emph{Phys. Rev.} {\bf
  D59} (1999) 044033}, [\href{http://arxiv.org/abs/hep-th/9809035}{{\tt
  hep-th/9809035}}].

\bibitem{Chamblin:1998pz}
A.~Chamblin, R.~Emparan, C.~V. Johnson and R.~C. Myers, \emph{{Large N phases,
  gravitational instantons and the nuts and bolts of AdS holography}},
  \href{http://dx.doi.org/10.1103/PhysRevD.59.064010}{\emph{Phys. Rev.} {\bf
  D59} (1999) 064010}, [\href{http://arxiv.org/abs/hep-th/9808177}{{\tt
  hep-th/9808177}}].

\bibitem{Hartnoll:2005yc}
S.~A. Hartnoll and S.~P. Kumar, \emph{{The O(N) model on a squashed S**3 and
  the Klebanov-Polyakov correspondence}},
  \href{http://dx.doi.org/10.1088/1126-6708/2005/06/012}{\emph{JHEP} {\bf 06}
  (2005) 012}, [\href{http://arxiv.org/abs/hep-th/0503238}{{\tt
  hep-th/0503238}}].

\bibitem{Hama:2011ea}
N.~Hama, K.~Hosomichi and S.~Lee, \emph{{SUSY Gauge Theories on Squashed
  Three-Spheres}}, \href{http://dx.doi.org/10.1007/JHEP05(2011)014}{\emph{JHEP}
  {\bf 05} (2011) 014}, [\href{http://arxiv.org/abs/1102.4716}{{\tt
  1102.4716}}].

\bibitem{Imamura:2011wg}
Y.~Imamura and D.~Yokoyama, \emph{{N=2 supersymmetric theories on squashed
  three-sphere}},
  \href{http://dx.doi.org/10.1103/PhysRevD.85.025015}{\emph{Phys. Rev.} {\bf
  D85} (2012) 025015}, [\href{http://arxiv.org/abs/1109.4734}{{\tt
  1109.4734}}].

\bibitem{Closset:2012ru}
C.~Closset, T.~T. Dumitrescu, G.~Festuccia and Z.~Komargodski,
  \emph{{Supersymmetric Field Theories on Three-Manifolds}},
  \href{http://dx.doi.org/10.1007/JHEP05(2013)017}{\emph{JHEP} {\bf 05} (2013)
  017}, [\href{http://arxiv.org/abs/1212.3388}{{\tt 1212.3388}}].

\bibitem{Jafferis:2011zi}
D.~L. Jafferis, I.~R. Klebanov, S.~S. Pufu and B.~R. Safdi, \emph{{Towards the
  F-Theorem: N=2 Field Theories on the Three-Sphere}},
  \href{http://dx.doi.org/10.1007/JHEP06(2011)102}{\emph{JHEP} {\bf 06} (2011)
  102}, [\href{http://arxiv.org/abs/1103.1181}{{\tt 1103.1181}}].

\bibitem{Perlmutter:2013gua}
E.~Perlmutter, \emph{{A universal feature of CFT Renyi entropy}},
  \href{http://dx.doi.org/10.1007/JHEP03(2014)117}{\emph{JHEP} {\bf 03} (2014)
  117}, [\href{http://arxiv.org/abs/1308.1083}{{\tt 1308.1083}}].

\bibitem{Cardy:1988cwa}
J.~L. Cardy, \emph{{Is There a c Theorem in Four-Dimensions?}},
  \href{http://dx.doi.org/10.1016/0370-2693(88)90054-8}{\emph{Phys. Lett.} {\bf
  B215} (1988) 749--752}.

\bibitem{Dowker:1998pi}
J.~S. Dowker, \emph{{Effective actions on the squashed three sphere}},
  \href{http://dx.doi.org/10.1088/0264-9381/16/6/323}{\emph{Class. Quant.
  Grav.} {\bf 16} (1999) 1937--1953},
  [\href{http://arxiv.org/abs/hep-th/9812202}{{\tt hep-th/9812202}}].

\bibitem{DeFrancia:2000xm}
M.~De~Francia, K.~Kirsten and J.~S. Dowker, \emph{{Effective actions on
  squashed lens spaces}},
  \href{http://dx.doi.org/10.1088/0264-9381/18/6/301}{\emph{Class. Quant.
  Grav.} {\bf 18} (2001) 955--968},
  [\href{http://arxiv.org/abs/hep-th/0008059}{{\tt hep-th/0008059}}].

\bibitem{Gibbons198098}
G.~Gibbons, \emph{Spectral asymmetry and quantum field theory in curved
  spacetime},
  \href{http://dx.doi.org/http://dx.doi.org/10.1016/0003-4916(80)90120-7}{\emph{Annals
  of Physics} {\bf 125} (1980) 98 -- 116}.

\bibitem{HITCHIN19741}
N.~Hitchin, \emph{Harmonic spinors},
  \href{http://dx.doi.org/http://dx.doi.org/10.1016/0001-8708(74)90021-8}{\emph{Advances
  in Mathematics} {\bf 14} (1974) 1 -- 55}.

\bibitem{Aharony:2008ug}
O.~Aharony, O.~Bergman, D.~L. Jafferis and J.~Maldacena, \emph{{N=6
  superconformal Chern-Simons-matter theories, M2-branes and their gravity
  duals}}, \href{http://dx.doi.org/10.1088/1126-6708/2008/10/091}{\emph{JHEP}
  {\bf 10} (2008) 091}, [\href{http://arxiv.org/abs/0806.1218}{{\tt
  0806.1218}}].

\bibitem{Emparan:1999pm}
R.~Emparan, C.~V. Johnson and R.~C. Myers, \emph{{Surface terms as counterterms
  in the AdS / CFT correspondence}},
  \href{http://dx.doi.org/10.1103/PhysRevD.60.104001}{\emph{Phys. Rev.} {\bf
  D60} (1999) 104001}, [\href{http://arxiv.org/abs/hep-th/9903238}{{\tt
  hep-th/9903238}}].

\bibitem{Mann:1999pc}
R.~B. Mann, \emph{{Misner string entropy}},
  \href{http://dx.doi.org/10.1103/PhysRevD.60.104047}{\emph{Phys. Rev.} {\bf
  D60} (1999) 104047}, [\href{http://arxiv.org/abs/hep-th/9903229}{{\tt
  hep-th/9903229}}].

\bibitem{Bueno1}
P.~Bueno, R.~C. Myers and W.~Witczak-Krempa, \emph{{Universality of corner
  entanglement in conformal field theories}},
  \href{http://dx.doi.org/10.1103/PhysRevLett.115.021602}{\emph{Phys. Rev.
  Lett.} {\bf 115} (2015) 021602}, [\href{http://arxiv.org/abs/1505.04804}{{\tt
  1505.04804}}].

\bibitem{Bueno2}
P.~Bueno and R.~C. Myers, \emph{{Corner contributions to holographic
  entanglement entropy}},
  \href{http://dx.doi.org/10.1007/JHEP08(2015)068}{\emph{JHEP} {\bf 08} (2015)
  068}, [\href{http://arxiv.org/abs/1505.07842}{{\tt 1505.07842}}].

\bibitem{Miao2015}
R.-X. Miao, \emph{{A holographic proof of the universality of corner
  entanglement for CFTs}},
  \href{http://dx.doi.org/10.1007/JHEP10(2015)038}{\emph{JHEP} {\bf 10} (2015)
  038}, [\href{http://arxiv.org/abs/1507.06283}{{\tt 1507.06283}}].

\bibitem{Bueno4}
P.~Bueno and R.~C. Myers, \emph{{Universal entanglement for higher dimensional
  cones}}, \href{http://dx.doi.org/10.1007/JHEP12(2015)168}{\emph{JHEP} {\bf
  12} (2015) 168}, [\href{http://arxiv.org/abs/1508.00587}{{\tt 1508.00587}}].

\bibitem{faulkner15}
T.~Faulkner, R.~G. Leigh and O.~Parrikar, \emph{{Shape Dependence of
  Entanglement Entropy in Conformal Field Theories}},
  \href{http://dx.doi.org/10.1007/JHEP04(2016)088}{\emph{JHEP} {\bf 04} (2016)
  088}, [\href{http://arxiv.org/abs/1511.05179}{{\tt 1511.05179}}].

\bibitem{Erdmenger:1996yc}
J.~Erdmenger and H.~Osborn, \emph{{Conserved currents and the energy momentum
  tensor in conformally invariant theories for general dimensions}},
  \href{http://dx.doi.org/10.1016/S0550-3213(96)00545-7}{\emph{Nucl. Phys.}
  {\bf B483} (1997) 431--474}, [\href{http://arxiv.org/abs/hep-th/9605009}{{\tt
  hep-th/9605009}}].

\bibitem{Hofman:2008ar}
D.~M. Hofman and J.~Maldacena, \emph{{Conformal collider physics: Energy and
  charge correlations}},
  \href{http://dx.doi.org/10.1088/1126-6708/2008/05/012}{\emph{JHEP} {\bf 05}
  (2008) 012}, [\href{http://arxiv.org/abs/0803.1467}{{\tt 0803.1467}}].

\bibitem{Lovelock2}
D.~Lovelock, \emph{{The Einstein tensor and its generalizations}},
  \href{http://dx.doi.org/10.1063/1.1665613}{\emph{J. Math. Phys.} {\bf 12}
  (1971) 498--501}.

\bibitem{Awad:2000gg}
A.~Awad and A.~Chamblin, \emph{{A Bestiary of higher dimensional Taub - NUT AdS
  space-times}},
  \href{http://dx.doi.org/10.1088/0264-9381/19/8/301}{\emph{Class. Quant.
  Grav.} {\bf 19} (2002) 2051--2062},
  [\href{http://arxiv.org/abs/hep-th/0012240}{{\tt hep-th/0012240}}].

\bibitem{Dehghani:2005zm}
M.~H. Dehghani and R.~B. Mann, \emph{{NUT-charged black holes in Gauss-Bonnet
  gravity}}, \href{http://dx.doi.org/10.1103/PhysRevD.72.124006}{\emph{Phys.
  Rev.} {\bf D72} (2005) 124006},
  [\href{http://arxiv.org/abs/hep-th/0510083}{{\tt hep-th/0510083}}].

\bibitem{Clarkson:2002uj}
R.~Clarkson, L.~Fatibene and R.~B. Mann, \emph{{Thermodynamics of
  (d+1)-dimensional NUT charged AdS space-times}},
  \href{http://dx.doi.org/10.1016/S0550-3213(02)01143-4}{\emph{Nucl. Phys.}
  {\bf B652} (2003) 348--382}, [\href{http://arxiv.org/abs/hep-th/0210280}{{\tt
  hep-th/0210280}}].

\bibitem{KhodamMohammadi:2008fh}
A.~Khodam-Mohammadi and M.~Monshizadeh, \emph{{Thermodynamics of
  Taub-NUT/Bolt-AdS Black Holes in Einstein-Gauss-Bonnet Gravity}},
  \href{http://dx.doi.org/10.1103/PhysRevD.79.044002}{\emph{Phys. Rev.} {\bf
  D79} (2009) 044002}, [\href{http://arxiv.org/abs/0811.1268}{{\tt
  0811.1268}}].

\bibitem{Buchel:2009tt}
A.~Buchel and R.~C. Myers, \emph{{Causality of Holographic Hydrodynamics}},
  \href{http://dx.doi.org/10.1088/1126-6708/2009/08/016}{\emph{JHEP} {\bf 08}
  (2009) 016}, [\href{http://arxiv.org/abs/0906.2922}{{\tt 0906.2922}}].

\bibitem{deBoer:2009pn}
J.~de~Boer, M.~Kulaxizi and A.~Parnachev, \emph{{AdS(7)/CFT(6), Gauss-Bonnet
  Gravity, and Viscosity Bound}},
  \href{http://dx.doi.org/10.1007/JHEP03(2010)087}{\emph{JHEP} {\bf 03} (2010)
  087}, [\href{http://arxiv.org/abs/0910.5347}{{\tt 0910.5347}}].

\bibitem{Ge:2009eh}
X.-H. Ge and S.-J. Sin, \emph{{Shear viscosity, instability and the upper bound
  of the Gauss-Bonnet coupling constant}},
  \href{http://dx.doi.org/10.1088/1126-6708/2009/05/051}{\emph{JHEP} {\bf 05}
  (2009) 051}, [\href{http://arxiv.org/abs/0903.2527}{{\tt 0903.2527}}].

\bibitem{Camanho:2009vw}
X.~O. Camanho and J.~D. Edelstein, \emph{{Causality constraints in AdS/CFT from
  conformal collider physics and Gauss-Bonnet gravity}},
  \href{http://dx.doi.org/10.1007/JHEP04(2010)007}{\emph{JHEP} {\bf 04} (2010)
  007}, [\href{http://arxiv.org/abs/0911.3160}{{\tt 0911.3160}}].

\bibitem{Banerjee:2014oaa}
S.~Banerjee, A.~Bhattacharyya, A.~Kaviraj, K.~Sen and A.~Sinha,
  \emph{{Constraining gravity using entanglement in AdS/CFT}},
  \href{http://dx.doi.org/10.1007/JHEP05(2014)029}{\emph{JHEP} {\bf 05} (2014)
  029}, [\href{http://arxiv.org/abs/1401.5089}{{\tt 1401.5089}}].

\bibitem{Brigante:2008gz}
M.~Brigante, H.~Liu, R.~C. Myers, S.~Shenker and S.~Yaida, \emph{{The Viscosity
  Bound and Causality Violation}},
  \href{http://dx.doi.org/10.1103/PhysRevLett.100.191601}{\emph{Phys. Rev.
  Lett.} {\bf 100} (2008) 191601}, [\href{http://arxiv.org/abs/0802.3318}{{\tt
  0802.3318}}].

\bibitem{Nishioka:2013gza}
T.~Nishioka and K.~Yonekura, \emph{{On RG Flow of $tau_{RR}$ for Supersymmetric
  Field Theories in Three-Dimensions}},
  \href{http://dx.doi.org/10.1007/JHEP05(2013)165}{\emph{JHEP} {\bf 05} (2013)
  165}, [\href{http://arxiv.org/abs/1303.1522}{{\tt 1303.1522}}].

\bibitem{Witczak-Krempa:2015jca}
W.~Witczak-Krempa and J.~Maciejko, \emph{{Optical conductivity of topological
  surface states with emergent supersymmetry}},
  \href{http://dx.doi.org/10.1103/PhysRevLett.117.149903,
  10.1103/PhysRevLett.116.100402}{\emph{Phys. Rev. Lett.} {\bf 116} (2016)
  100402}, [\href{http://arxiv.org/abs/1510.06397}{{\tt 1510.06397}}].

\bibitem{Martelli:2011fu}
D.~Martelli, A.~Passias and J.~Sparks, \emph{{The gravity dual of
  supersymmetric gauge theories on a squashed three-sphere}},
  \href{http://dx.doi.org/10.1016/j.nuclphysb.2012.07.019}{\emph{Nucl. Phys.}
  {\bf B864} (2012) 840--868}, [\href{http://arxiv.org/abs/1110.6400}{{\tt
  1110.6400}}].

\bibitem{PhysRevD.8.1048}
B.~L. Hu, \emph{{Scalar Waves in the Mixmaster Universe. I. The Helmholtz
  Equation in a Fixed Background}},
  \href{http://dx.doi.org/10.1103/PhysRevD.8.1048}{\emph{Phys. Rev. D} {\bf 8}
  (Aug, 1973) 1048--1060}.

\bibitem{Casini:2011kv}
H.~Casini, M.~Huerta and R.~C. Myers, \emph{{Towards a derivation of
  holographic entanglement entropy}},
  \href{http://dx.doi.org/10.1007/JHEP05(2011)036}{\emph{JHEP} {\bf 05} (2011)
  036}, [\href{http://arxiv.org/abs/1102.0440}{{\tt 1102.0440}}].

\bibitem{Dowker:2010yj}
J.~S. Dowker, \emph{{Entanglement entropy for odd spheres}},
  \href{http://arxiv.org/abs/1012.1548}{{\tt 1012.1548}}.

\bibitem{Allais:2014ata}
A.~Allais and M.~Mezei, \emph{{Some results on the shape dependence of
  entanglement and Renyi entropies}},
  \href{http://dx.doi.org/10.1103/PhysRevD.91.046002}{\emph{Phys. Rev.} {\bf
  D91} (2015) 046002}, [\href{http://arxiv.org/abs/1407.7249}{{\tt
  1407.7249}}].

\bibitem{Mezei14}
M.~Mezei, \emph{{Entanglement entropy across a deformed sphere}},
  \href{http://dx.doi.org/10.1103/PhysRevD.91.045038}{\emph{Phys. Rev.} {\bf
  D91} (2015) 045038}, [\href{http://arxiv.org/abs/1411.7011}{{\tt
  1411.7011}}].

\bibitem{Rosenhaus:2014zza}
V.~Rosenhaus and M.~Smolkin, \emph{{Entanglement Entropy for Relevant and
  Geometric Perturbations}},
  \href{http://dx.doi.org/10.1007/JHEP02(2015)015}{\emph{JHEP} {\bf 02} (2015)
  015}, [\href{http://arxiv.org/abs/1410.6530}{{\tt 1410.6530}}].

\bibitem{Rosenhaus:2014woa}
V.~Rosenhaus and M.~Smolkin, \emph{{Entanglement Entropy: A Perturbative
  Calculation}}, \href{http://dx.doi.org/10.1007/JHEP12(2014)179}{\emph{JHEP}
  {\bf 12} (2014) 179}, [\href{http://arxiv.org/abs/1403.3733}{{\tt
  1403.3733}}].

\bibitem{Lewkowycz:2014jia}
A.~Lewkowycz and E.~Perlmutter, \emph{{Universality in the geometric dependence
  of Renyi entropy}},
  \href{http://dx.doi.org/10.1007/JHEP01(2015)080}{\emph{JHEP} {\bf 01} (2015)
  080}, [\href{http://arxiv.org/abs/1407.8171}{{\tt 1407.8171}}].

\bibitem{Bianchi:2015liz}
L.~Bianchi, M.~Meineri, R.~C. Myers and M.~Smolkin, \emph{{R\'enyi entropy and
  conformal defects}},
  \href{http://dx.doi.org/10.1007/JHEP07(2016)076}{\emph{JHEP} {\bf 07} (2016)
  076}, [\href{http://arxiv.org/abs/1511.06713}{{\tt 1511.06713}}].

\bibitem{Bueno5}
P.~Bueno and W.~Witczak-Krempa, \emph{{Bounds on corner entanglement in quantum
  critical states}},
  \href{http://dx.doi.org/10.1103/PhysRevB.93.045131}{\emph{Phys. Rev.} {\bf
  B93} (2016) 045131}, [\href{http://arxiv.org/abs/1511.04077}{{\tt
  1511.04077}}].

\bibitem{Carmi:2015dla}
D.~Carmi, \emph{{On the Shape Dependence of Entanglement Entropy}},
  \href{http://dx.doi.org/10.1007/JHEP12(2015)043}{\emph{JHEP} {\bf 12} (2015)
  043}, [\href{http://arxiv.org/abs/1506.07528}{{\tt 1506.07528}}].

\bibitem{Dong:2016wcf}
X.~Dong, \emph{{Shape Dependence of Holographic RŽnyi Entropy in Conformal
  Field Theories}},
  \href{http://dx.doi.org/10.1103/PhysRevLett.116.251602}{\emph{Phys. Rev.
  Lett.} {\bf 116} (2016) 251602}, [\href{http://arxiv.org/abs/1602.08493}{{\tt
  1602.08493}}].

\bibitem{Bianchi:2016xvf}
L.~Bianchi, S.~Chapman, X.~Dong, D.~A. Galante, M.~Meineri and R.~C. Myers,
  \emph{{Shape dependence of holographic RŽnyi entropy in general dimensions}},
  \href{http://dx.doi.org/10.1007/JHEP11(2016)180}{\emph{JHEP} {\bf 11} (2016)
  180}, [\href{http://arxiv.org/abs/1607.07418}{{\tt 1607.07418}}].

\bibitem{Balakrishnan:2016ttg}
S.~Balakrishnan, S.~Dutta and T.~Faulkner, \emph{{Gravitational dual of the
  R\'{e}nyi twist displacement operator}},
  \href{http://arxiv.org/abs/1607.06155}{{\tt 1607.06155}}.

\bibitem{Fonda:2015nma}
P.~Fonda, D.~Seminara and E.~Tonni, \emph{{On shape dependence of holographic
  entanglement entropy in AdS$_{4}$/CFT$_{3}$}},
  \href{http://dx.doi.org/10.1007/JHEP12(2015)037}{\emph{JHEP} {\bf 12} (2015)
  037}, [\href{http://arxiv.org/abs/1510.03664}{{\tt 1510.03664}}].

\bibitem{Gibbons:1979xm}
G.~W. Gibbons and S.~W. Hawking, \emph{{Classification of Gravitational
  Instanton Symmetries}},
  \href{http://dx.doi.org/10.1007/BF01197189}{\emph{Commun. Math. Phys.} {\bf
  66} (1979) 291--310}.

\bibitem{Gibbons:1976ue}
G.~W. Gibbons and S.~W. Hawking, \emph{{Action Integrals and Partition
  Functions in Quantum Gravity}},
  \href{http://dx.doi.org/10.1103/PhysRevD.15.2752}{\emph{Phys. Rev.} {\bf D15}
  (1977) 2752--2756}.

\bibitem{Myers:1987yn}
R.~C. Myers, \emph{{Higher Derivative Gravity, Surface Terms and String
  Theory}}, \href{http://dx.doi.org/10.1103/PhysRevD.36.392}{\emph{Phys. Rev.}
  {\bf D36} (1987) 392}.

\bibitem{Teitelboim:1987zz}
C.~Teitelboim and J.~Zanelli, \emph{{Dimensionally continued topological
  gravitation theory in Hamiltonian form}},
  \href{http://dx.doi.org/10.1088/0264-9381/4/4/010}{\emph{Class. Quant. Grav.}
  {\bf 4} (1987) L125}.

\bibitem{Skenderis:2002wp}
K.~Skenderis, \emph{{Lecture notes on holographic renormalization}},
  \href{http://dx.doi.org/10.1088/0264-9381/19/22/306}{\emph{Class. Quant.
  Grav.} {\bf 19} (2002) 5849--5876},
  [\href{http://arxiv.org/abs/hep-th/0209067}{{\tt hep-th/0209067}}].

\bibitem{Balasubramanian:1999re}
V.~Balasubramanian and P.~Kraus, \emph{{A Stress tensor for Anti-de Sitter
  gravity}}, \href{http://dx.doi.org/10.1007/s002200050764}{\emph{Commun. Math.
  Phys.} {\bf 208} (1999) 413--428},
  [\href{http://arxiv.org/abs/hep-th/9902121}{{\tt hep-th/9902121}}].

\bibitem{Brihaye:2008xu}
Y.~Brihaye and E.~Radu, \emph{{Black objects in the Einstein-Gauss-Bonnet
  theory with negative cosmological constant and the boundary counterterm
  method}}, \href{http://dx.doi.org/10.1088/1126-6708/2008/09/006}{\emph{JHEP}
  {\bf 09} (2008) 006}, [\href{http://arxiv.org/abs/0806.1396}{{\tt
  0806.1396}}].

\bibitem{Slansky:1981yr}
R.~Slansky, \emph{{Group Theory for Unified Model Building}},
  \href{http://dx.doi.org/10.1016/0370-1573(81)90092-2}{\emph{Phys. Rept.} {\bf
  79} (1981) 1--128}.

\bibitem{Vassilevich:2003xt}
D.~V. Vassilevich, \emph{{Heat kernel expansion: User's manual}},
  \href{http://dx.doi.org/10.1016/j.physrep.2003.09.002}{\emph{Phys. Rept.}
  {\bf 388} (2003) 279--360}, [\href{http://arxiv.org/abs/hep-th/0306138}{{\tt
  hep-th/0306138}}].

\end{thebibliography}\endgroup
\label{biblio}
  
\end{document}